\documentclass{IEEEaccess}
\usepackage{cite}
\usepackage{amsmath,amssymb,amsfonts}
\usepackage{graphicx}
\usepackage{textcomp}

\usepackage{algpseudocode}
\usepackage{algorithm}
\usepackage{multirow}
\usepackage{array}
\usepackage{booktabs}
\usepackage{bm}

\usepackage{soul}
\usepackage[utf8]{inputenc}
\usepackage{float}
\usepackage{cases}
\usepackage{mdwmath}

\usepackage{colortbl}
\usepackage[normalem]{ulem}
\usepackage{siunitx}
\usepackage{hyperref}
\makeatletter
\let\MYcaption\@makecaption
\makeatother

\usepackage[font=footnotesize]{subcaption}

\makeatletter
\let\@makecaption\MYcaption
\makeatother

\def\BibTeX{{\rm B\kern-.05em{\sc i\kern-.025em b}\kern-.08em
    T\kern-.1667em\lower.7ex\hbox{E}\kern-.125emX}}
\begin{document}
\history{This article has been accepted for publication in a future issue of this journal, but has not been fully edited. Content may change prior to final publication. Citation information: M. Ahishali \textit{et al.}, "Advance Warning Methodologies for COVID-19 Using Chest X-Ray Images," in \textit{IEEE Access}, vol. 9, pp. 41052-41065, 2021.}
\doi{10.1109/ACCESS.2021.3064927}

\title{Advance Warning Methodologies for COVID-19 using Chest X-Ray Images}
\author{\uppercase{Mete Ahishali}\authorrefmark{1},
\uppercase{Aysen Degerli}\authorrefmark{1}, 
\uppercase{Mehmet Yamac}\authorrefmark{1}, 
\uppercase{Serkan Kiranyaz}\authorrefmark{2}, \IEEEmembership{Senior Member, IEEE}, 
\uppercase{Muhammad E. H. Chowdhury}\authorrefmark{2}, \IEEEmembership{Senior Member, IEEE},
\uppercase{Khalid Hameed}\authorrefmark{3}, 
\uppercase{Tahir Hamid}\authorrefmark{4, 5}, 
\uppercase{Rashid Mazhar\authorrefmark{4}, Moncef Gabbouj}\authorrefmark{1}, \IEEEmembership{Fellow, IEEE}.}
\address[1]{Faculty of Information Technology and Communication Sciences, Tampere University, 33720 Tampere, Finland (email: name.surname@tuni.fi)}
\address[2]{Department of Electrical Engineering, Qatar University, 2713 Doha, Qatar (email: mkiranyaz@qu.edu.qa and mchowdhury@qu.edu.qa)}
\address[3]{Reem Medical Center, Doha, Qatar (email: dr.khalid@reemmedicalcenter.com)}
\address[4]{Hamad Medical Corporation Hospital, Doha, Qatar}
\address[5]{Weill Cornell Medicine - Qatar, Doha, Qatar}
\tfootnote{This work was supported by the Academy of Finland's project AWcHA, Dn 334566.}

\markboth
{Ahishali \headeretal: Advance Warning Methodologies for COVID-19 using Chest X-Ray Images}
{Ahishali \headeretal:Advance Warning Methodologies for COVID-19 using Chest X-Ray Images}

\corresp{Corresponding author: Mete Ahishali (e-mail: mete.ahishali@tuni.fi).}

\begin{abstract}
Coronavirus disease 2019 (COVID-19) has rapidly become a global health concern after its first known detection in December 2019. As a result, accurate and reliable advance warning system for the early diagnosis of COVID-19 has now become a priority. The detection of COVID-19 in early stages is not a straightforward task from chest X-ray images according to expert medical doctors because the traces of the infection are visible only when the disease has progressed to a moderate or severe stage. In this study, our first aim is to evaluate the ability of recent \textit{state-of-the-art} Machine Learning techniques for the early detection of COVID-19 from chest X-ray images. Both compact classifiers and deep learning approaches are considered in this study. Furthermore, we propose a recent compact classifier, Convolutional Support Estimator Network (CSEN) approach for this purpose since it is well-suited for a scarce-data classification task. Finally, this study introduces a new benchmark dataset called Early-QaTa-COV19\footnote{The Early-QaTa-COV19 dataset is publicly shared at the repository \href{https://www.kaggle.com/aysendegerli/qatacov19-dataset}{https://www.kaggle.com/aysendegerli/qatacov19-dataset}.}, which consists of 1065 early-stage COVID-19 pneumonia samples (very limited or no infection signs) labelled by the medical doctors and 12 544 samples for control (normal) class. A detailed set of experiments shows that the CSEN achieves the top (over 97\%) sensitivity with over 95.5\% specificity \footnote{The software implementations of the methods are available at \href{https://github.com/meteahishali/methods-early-cov19}{https://github.com/meteahishali/methods-early-cov19}.}. Moreover, DenseNet-121 network produces the leading performance among other deep networks with 95\% sensitivity and 99.74\% specificity.
\end{abstract}


\titlepgskip=-15pt

\maketitle

\section{Introduction}
\label{sec:introduction}
\PARstart{C}{oronavirus} disease 2019 (COVID-19) caused by severe acute respiratory syndrome coronavirus-2 (SARS-CoV-2) has declared as pandemic by the World Health Organization (WHO) on the 11th of March, 2020. Although, infected patients tend to have mild and unspecific symptoms \cite{cov-19_2} such as fever, myalgia or fatigue, and cough, the disease affects seriously people in high-risk groups especially the elderly. Up to now, COVID-19 has caused more than one-million fatalities.

\begin{figure*}[t]
\centering
\includegraphics[width=0.98\textwidth]{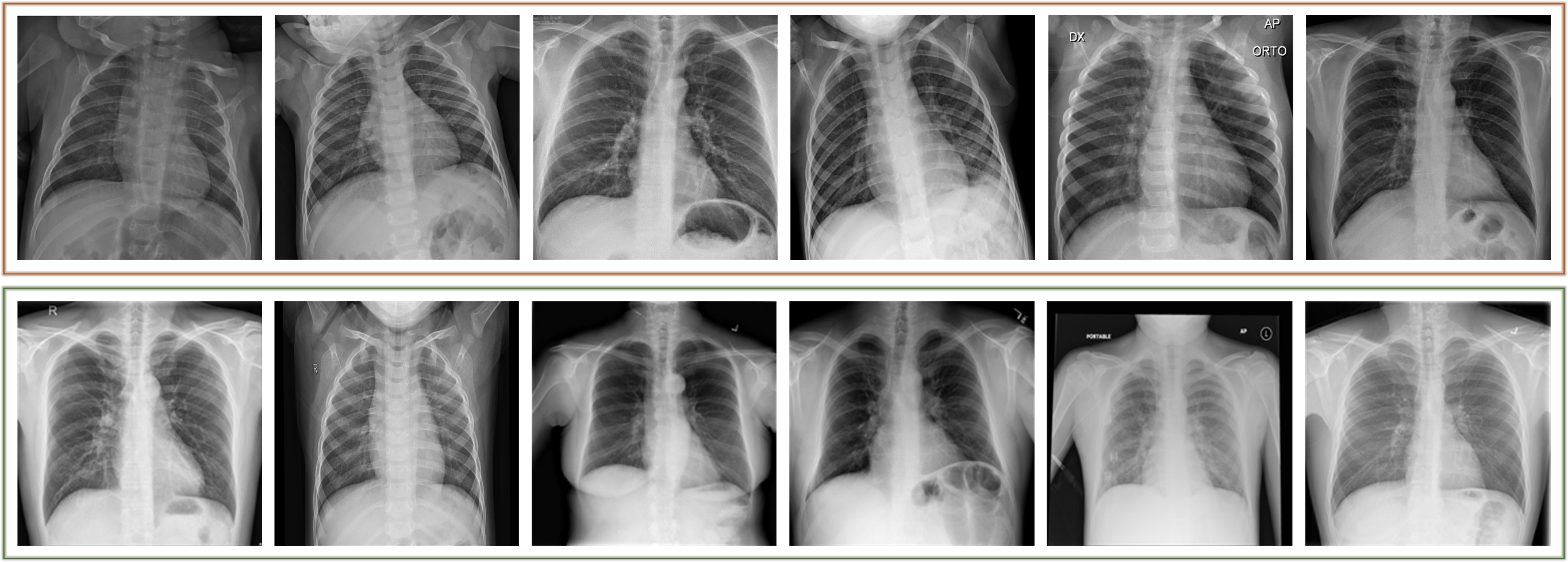}
\caption{(first row) Samples of COVID-19 pneumonia with very limited or no visible sign of COVID-19, and (second row) normal (healthy) class from Early-QaTa-COV19 dataset.}
\label{fig:interclass}
\end{figure*}

There have been different detection methods for COVID-19. In clinics, reverse transcription polymerase chain reaction (RT-PCR) has been used and it holds the reference method \cite{cov-19_2} for COVID-19 detection. It is also recommended by WHO that a rapid collection of suitable specimens from suspect cases should be made with RT-PCR like nucleic acid amplification tests \cite{world2020laboratory} and implies the vital role of RT-PCR to prevent spreading of the disease. However, RT-PCR is known to have a low sensitivity, and it is reported in \cite{pcr_positive_rate} that RT-PCR has around $30-60\%$ total positive rate for throat samples, and low positive rates occur especially in mild cases. To this end, there are studies \cite{cov-19, ct, sensitivity_ct} that investigate the usage of Chest-CTs and correlation between Chest-CT and RT-PCR tests as diagnostic tools. It is stated in \cite{cov-19} that Chest-CT scans have positive findings for $75\%$ of negative RT-PCR samples, and \cite{ct} suggests to repeat swap testing for the cases where CT scans have suspicious finding even though RT-PCR results are negative. Finally, \cite{sensitivity_ct} calculates the sensitivities of Chest-CT and RT-PCR as $98$ and $71\%$, respectively.

Although the above-mentioned studies propose to use Chest-CT scans in epicenters rather than RT-PCR to detect COVID-19 where RT-PCR has a low sensitivity for mild cases, there are several limitations of CT scans such as the time for image acquisition, the associated cost, and availability of CT devices. On the other hand, X-ray imaging is a highly available and faster diagnostic tool. Unlike CTs, X-ray imaging is also cheaper, and patients are less harmed from radiation \cite{ct_harm} during the acquisition process. Another advantage is that there are portable X-ray devices, and hence, as stated in \cite{xray_ad}, X-ray can reduce the risk of contamination compared to CT for suspects where the person can spread the disease in the transport route. Overall, chest X-ray images can be an alternative for COVID-19 detection with other diagnosis tools (for example, RT-PCR) especially in heavily affected areas where the detection delay is critical and the resources are limited.

The outbreak has brought the urgent need for an automated, accurate, and robust COVID-19 detection/recognition system that can guide the practitioner to diagnose suspects especially in early stages. For example, many countries suffer from incorrect statistics because of the time-consuming part of the manual diagnostic tools \cite{Xray1}. Although, there are existing studies for COVID-19 recognition using Chest-CT scans such as \cite{ct-detection-auto}, several studies \cite{Xray1, Xray2, Xray3, exact4, csen_covid} propose to use X-ray images for automated COVID-19 recognition because of the aforementioned advantages of X-ray imaging. A detailed survey study in \cite{review} reviews COVID-19 detection approaches using chest X-ray images and Chest-CT scans. Nevertheless, all of them except \cite{csen_covid} have been experimented over only a small amount of data, e.g., the largest one includes only a few hundreds of X-ray images with only few COVID-19 samples. To address this need, in an earlier study \cite{csen_covid}, we have compiled QaTa-COV19 dataset with $462$ chest X-ray images from COVID-19 patients. Then, COVID-19 samples are further populated to 2951 images in \cite{degerli2020covid} with their corresponding ground-truth segmentation masks indicating the infected regions on the lungs.

\begin{figure}[t]
\centering
\includegraphics[width=0.45\textwidth]{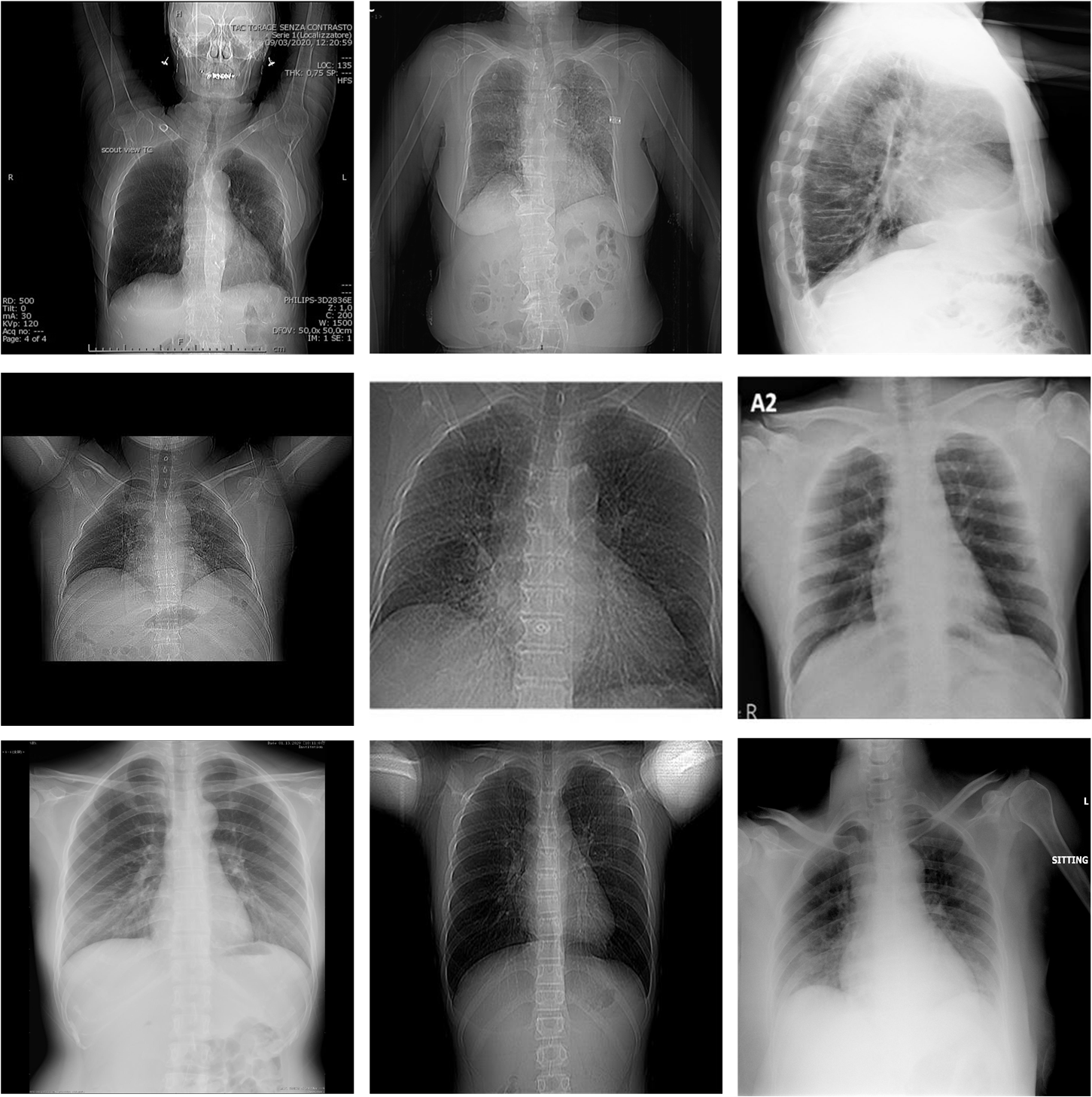}
\caption{Sample X-ray images from Early-QaTa-COV19 in different quality, resolution and noise level and showing no or very limited sign of COVID-19 pneumonia.}
\label{fig:intraclass}
\end{figure}

As stated in \cite{cov-19, cov-19_2}, early detection plays a vital role to prevent spreading the disease by detecting infected people, isolating them, starting the treatment, and preventing possible secondary infections on the same patient. In this study, the "early" term in the detection task is used in the following sense: the selected COVID-19 chest X-ray images do not have any Pneumonic changes on the lungs or they have very limited signs that can easily be misinterpreted by the medical doctors (MDs). Consequently, the early stage is not measured by the time, but rather with visibility of the infection traces. On one hand, COVID-19 detection from chest X-ray images is a straightforward task when it is already in late-stage and the patient's X-ray shows moderate or severe signs of infection; hence, it is well-studied in the literature. However, there is a lack of research on the automated early detection of COVID-19 using X-ray images. This can be due to the following reasons: during the early stage, the detection can be difficult or perhaps not feasible at all even for an expert MD using X-ray images. For example, in many studies \cite{cov-19_2, xray_ad, chest_early}, it is stated that chest X-ray images are not sensitive compared to CT scans for the early detection where the symptoms are mild, and they further claim that there can be traces of the infection that can only be detected by MDs in severe patients. Therefore, in this study, our first aim is to investigate state-of-the-art Machine Learning (ML) approaches for their usability in advance warning for COVID-19 using chest X-ray images. To accomplish this objective, we have first compiled a new benchmark dataset called Early-QaTa-COV19, which is formed from QaTa-COV19 with some additional images. For this purpose, X-ray images from the COVID-19 patients who are in the early stages of the disease are selected by a group of MDs. As some of them shown in Fig. \ref{fig:interclass}, these samples have limited or no visible sign of COVID-19 pneumonia observed by the human eye. Accordingly, the Early-QaTa-COV19 dataset consists of $1065$ COVID-19 samples ($801$ and $264$ images with no and limited infection signs, respectively) and $12 544$ samples from the control (normal) class. Early-QaTa-COV19 dataset has several unique properties: the first and foremost, it is extracted from the largest benchmark dataset, QaTa-COV19, ever formed in the World and further populated with new X-ray images. Next, the dataset is the most diverse database encapsulating X-ray images from numerous countries (e.g. Italy, Spain, China, etc.) and different X-ray machines. Consequently, the images are in different quality, resolution, and noise levels as shown in Fig. \ref{fig:intraclass}.

As a consequence of recent advances in Deep Learning, techniques that are based on Convolutional Neural Networks (CNNs) have achieved state-of-the-art performances in many computer vision tasks such as image segmentation, recognition, and object detection. However, there are certain limitations on the approaches based on deep CNNs: the requirement of a large dataset to achieve the required generalization capability of such deep networks, and the necessity of an additional graphical processing unit (GPU) hardware to achieve reasonable inference time. Nevertheless, there are existing studies addressing these drawbacks, for example, the method in \cite{wang2019saliencygan} is able to reduce the volume of the training data by $10\%$ using a semi-supervised based approach. Moreover, for the latter, the method in \cite{knowledgedistillation} proposes a knowledge distillation approach to transfer CNN models to source-limited devices where only light-weight models can be fit and suitable for the inference. Next, the need for such lightweight conversions is discussed in \cite{machineiot} for IoT devices.

Unlike deep learners, traditional supervised approaches can also be utilized especially when the data is scarce. For example, representation-based classification approaches consisting of Sparse Representation based Classification (SRC) \cite{SRC1, SRC2} and Collaborative Representation based Classification (CRC) \cite{collaborative} are proven to perform well with a limited data. Accordingly, in the representation-based classification approaches, a dictionary $\mathbf{D}$ is formed by stacking samples from the training set. Then, when a test sample $\textbf{y}$ is introduced, it is assumed that the query samples can be represented as a linear combination of the atoms in $\mathbf{D}$. Therefore, the estimated representation coefficients, $\mathbf{\hat{x}}$ that is obtained by solving $\mathbf{y} = \mathbf{D x}$, carry enough information about the class of $\mathbf{y}$. For example, SRC approaches compute sparse solutions: the estimated $\mathbf{\hat{x}}$ has just enough non-zero coefficients, where only corresponding samples with the same query class in the dictionary $\mathbf{D}$ contributes. As SRC approaches, \cite{SRC1, SRC2}, provide slightly improved results compared to CRC, they are iterative methods and computationally complex. On the other hand, CRC provides a non-iterative and a relatively faster alternative via least-square sense solution, yet it produces comparable results as presented in \cite{collaborative}.

\begin{figure*}[h!]
    \centering
    \includegraphics[width=0.99\textwidth]{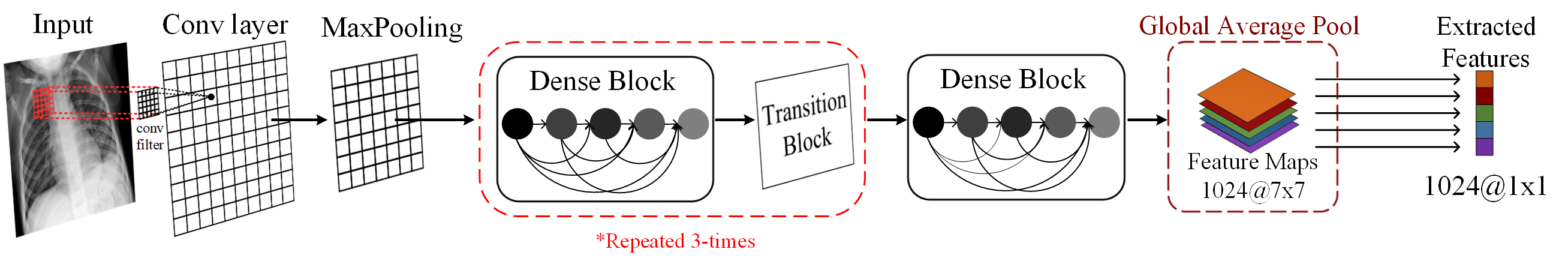}
    \caption{Feature extraction pipeline from the pre-trained DenseNet-121. 1024-D feature vectors are extracted for the compact classifiers trained for the early detection of COVID-19.}
    \label{fig:FX}
\end{figure*}

Convolutional Support Estimator Network (CSEN) introduced in a recent work \cite{CSEN} is proposed to combine traditional representation-based classification with a learning-based methodology. We define the support set as the location of non-zero elements in $\mathbf{x}$. Accordingly, the support set is more important than the exact values of $\mathbf{x}$ since it reveals the class information of the query. It was validated in \cite{CSEN} that CSENs provide a superior classification performance and computational efficiency against other representation-based methods by directly learning the mapping from query sample $\mathbf{y}$ to the corresponding support set from a small amount of training data. Moreover, CSENs are evaluated in our previous study \cite{csen_covid} for COVID-19 recognition in the benchmark QaTa-COV19 dataset where it has achieved over $98\%$ sensitivity and $95\%$ specificity for COVID-19 recognition. Therefore, with their capabilities of performing well with limited training dataset, both traditional representation-based classifiers and CSENs are good candidates for Early-QaTa-COV19 dataset.

Overall, in this study, we evaluate compact and deep classifiers for a possible advance warning system to detect early stages of COVID-19 from X-ray images using the Early-QaTa-COV19 dataset. The evaluated methods include compact and deep classifiers using the Early-QaTa-COV19 dataset. In the former group, SRC, CRC, CSEN, Multi-Layer Perceptron (MLP), Support Vector Machine (SVM), and k-Nearest Neighbor (k-NN) classifiers are evaluated. In the latter group, we evaluate the following deep networks: DenseNet-121 \cite{DenseNet}, ResNet-50 \cite{resnet50}, and Inception-v3 \cite{inception} networks.

In this manner, the contributions of this study can be summarized as the following:
\begin{itemize}
    \item We compile the world's largest dataset Early-QaTa-COV19 for early COVID-19 detection with chest X-ray images labelled by our group of MDs.
	\item This study investigates the feasibility of early detection and advance warning for COVID-19 from chest X-ray images, which exhibit limited or no sign of the infection by the naked eye. To accomplish this task, in this study, we perform an extensive set of comparative evaluations among numerous state-of-the-art techniques over the benchmark dataset compiled for this purpose.
	\item To this end, it is for the first time the CSEN approach is compared against several state-of-the-art approaches including deep CNNs by providing an extensive set of evaluations.
\end{itemize}

The results demonstrate that it is possible to achieve a robust and highly accurate early COVID-19 detection with a tolerable false alarm rate by CSENs using the deep features, whereas deep learners provide reduced sensitivity levels but with improved specificity.

The rest of the paper is organized as follows: a brief overview and preliminaries will be provided related to generic sparse representation in Section \ref{preliminaries}. Next, the state-of-the-art classification methods used in this study are detailed in Section \ref{methods}. The experimental results with the benchmark Early-QaTa-COV19 dataset are presented in Section \ref{experimental}. Finally, Section \ref{conclusion} concludes the paper.

\begin{figure*}[ht!]
    \centering
    \includegraphics[width=0.9\linewidth]{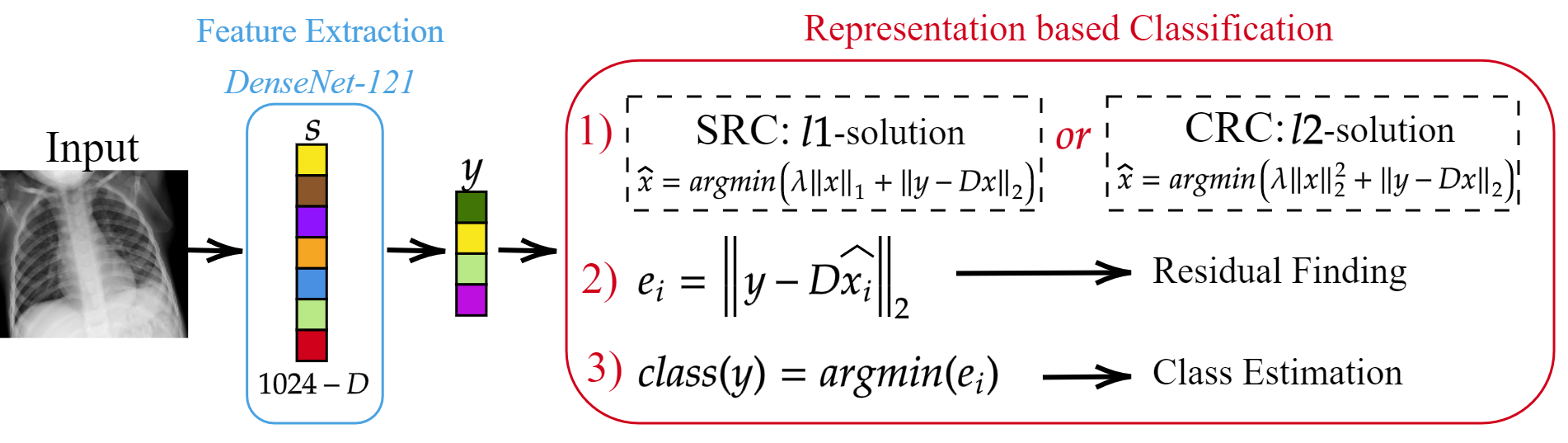}
    \caption{Representation-based classification pipeline for the early detection of COVID-19 using chest X-ray images.}
    \label{fig:src_crc}
\end{figure*}

\section{Background and Preliminaries}
\label{preliminaries}
In this section, a brief overview will be provided for the representation-based classification approaches used in this study. Accordingly, we define the following notations: $\left \| \mathbf{x} \right \|_{\ell_p^n} =   \left (  \sum_{i=1}^n \left \vert x_i \right \vert^p \right )^{1/p}$ is the $\ell_p$-norm of a vector $\mathbf{x} \in \mathbb{R}^n$ where $p \geq 1$. The $\ell_{\infty}$-norm and $\ell_0$-norm are defined for the vector $\mathbf{x}$ as $\left \| \mathbf{x} \right \|_{\ell_{\infty}^n} =  \max_{i=1,...,n} \left (   \left | x_i \right | \right )$ and $\left \| \mathbf{x} \right \|_{\ell_0^n} = \lim_{p \to 0} \sum_{i=1}^n \left \vert x_i \right \vert^p = \# \{ j: x_j \neq 0 \}$, respectively.

Additionally, a signal $\mathbf{s}$ can be called as strictly $k$-sparse if there is a proper domain $ \mathbf{ \Phi}$ that can represent the signal with less than $k+1$ non-zero coefficients: $ \mathbf{s}= \mathbf{ \Phi}\ \mathbf{x}$ where $\left \|  \mathbf{x} \right \|_0 \leq k$. In other words, the signal $\mathbf{s}$ can be represented in some domain, i.e., $ \mathbf{ \Phi}$, using only small number of basis vectors. Hence, we define a (sparse) support set of the signal $\mathbf{x}$, $\Lambda \subset \{1,2,3,...,n \}$, as $\Lambda := \left \{ i:  x_i =0 \right \}$ indexing the non-zero coefficients of $\mathbf{x}$ which corresponds these basis components.

The signal, $\mathbf{s}$, can be represented in a subspace $\mathbf{A}$, i.e., $\mathbf{y}=\mathbf{As}$. Accordingly, $\mathbf{y}$ can be sparse coded in the equivalent dictionary, $\mathbf{D}$, as follows:
\begin{equation}
     \mathbf{y} = \mathbf{A} \mathbf{s} = \mathbf{A}\mathbf{ \Phi} \mathbf{x} =\mathbf{ D} \mathbf{x} \label{CS},
\end{equation}
where $\mathbf{D} \in \mathbb{R}^{m \times n}$, and $\mathbf{A} \in \mathbb{R}^{m \times d}$ is the compression matrix for $m << d$. If the signal $\mathbf{x}$ is $k$-sparse in a $\mathbf{ \Phi}$ sparsifying basis, then, the solution of
\begin{equation}
\min_\mathbf{x} ~ \left \| \mathbf{x }\right \|_{0}~ \text{subject to}~ \mathbf{D} \mathbf{x} = \mathbf{y} \label{sparse_rep}
\end{equation}
is unique if $\left \| \mathbf{x }\right \|_{0} \leq k$ and $m \geq 2k$ \cite{spark}. Hence, it can be said based on \eqref{sparse_rep} that at least $k$-sparse signal pairs can be distinguishable in the equivalent dictionary, $\mathbf{D}$.

As the above-mentioned optimization problem is non-convex, and NP-hard, its relaxation by $\ell_1$ can be applied which is the closest convex norm:
\begin{equation}
     \min_\mathbf{x} \left \| \mathbf{x} \right \|_1 ~s.t. ~ \mathbf{x} \in \mho \left (\mathbf{ y} \right ) \label{Eq:l1}
\end{equation}
which is defined as Basis Pursuit \cite{BP} where $\mho \left ( \mathbf{y} \right ) = \left \{ \mathbf{x}: \mathbf{D} \mathbf{x}=\mathbf{y} \right \}$.

On the other hand, as previously discussed, in representation-based classification \cite{SRC2,collaborative,CSEN}, estimating the support set, $\Lambda$, would be more beneficial than the signal recovery. Let the support estimator be $\mathcal{E}(.,.)$ for a linear measurement scheme with an additive noise $\mathbf{y}= \mathbf{Dx} +\mathbf{z}$:
\begin{equation}
   \hat{\Lambda} =  \mathcal{E}\left (\mathbf{y},\mathbf{D} \right ) 
\end{equation}
where $\hat{\Lambda}$ is the estimation.

In practice, the performance of the recovery of $\Lambda$ is related with the recovery performance of the sparse signal in traditional SE methods, since they are based on first applying a signal recovery method, then, applying component-wise thresholding over the estimated signal, $\hat{\mathbf{x}}$, to compute $\hat{\Lambda}$. However, in \cite{CSEN}, we have shown that the direct recovery most likely causes noisy estimation while CSEN is able to learn sparse patterns and accomplish better SE compared to the competing methods. The readers are referred to \cite{CSEN} for a more detailed survey and evaluations on the support estimation performances indicating the limitations and drawbacks of traditional methods compared to the proposed CSEN approach.

\section{Methods for Early Detection of COVID-19}
\label{methods}

\subsection{Compact Approaches for Early Detection}
In this section, we present the state-of-the-art methodologies and explain our configurations for their application to early detection of COVID-19. First, we present a feature extraction procedure along with the compact classifier approaches in the first group, then a detailed discussion is provided on the chosen deep networks for the early detection problem. Note the fact that the methods in the first group are selected considering their suitability for the early detection task where the training data is scarce.

\subsubsection{Feature extraction by DenseNet-121} \label{feature_extraction}
Traditional ML approaches need feature extraction for classification. In accordance with the purpose of this study, we consider the pre-trained DenseNet-121 models for the feature extraction trained on Early-QaTa-COV19 and ChestX-ray14 datasets \cite{XrayDataset}. The ChestX-ray14 dataset consists of $14$ different pathology classes. The provided pre-trained network in \cite{chexnet} is initialized with ImageNet weights, and it is fine-tuned over 100 000 chest X-ray images. It is reported in \cite{chexnet} that DenseNet-121 produces the best results on the ChestX-ray14 dataset, and it also achieves better performance levels than radiologists' average decisions.

In this study, the pre-trained two DenseNet-121 models on Early-QaTa-COV19 and ChestX-ray14 datasets are used to extract 1024-D feature vectors by taking the output after global pooling just before the classification layer, which is illustrated in Fig. \ref{fig:FX}. Then, a dimensionality reduction is applied over the calculated features with principal component analysis (PCA) by choosing the first $512$ principal components. Hence, for a feature vector $\mathbf{s} \in \mathbb{R}^{d=1024}$, the query sample $\mathbf{y} = \mathbf{A} \mathbf{s} \in  \mathbb{R}^m$ is computed, where $\mathbf{A} \in \mathbb{R}^{m \times d} $ is PCA matrix computed over the training data and $m<d$. Then, data normalization is applied over the calculated $\mathbf{y}$ to have zero mean and unit variance for MLP, SVM, and k-NN classifiers, and zero mean and unit-norm for SRC, CRC, and CSEN approaches.

\begin{figure*}[t!]
    \centering
    \includegraphics[width=0.9\linewidth]{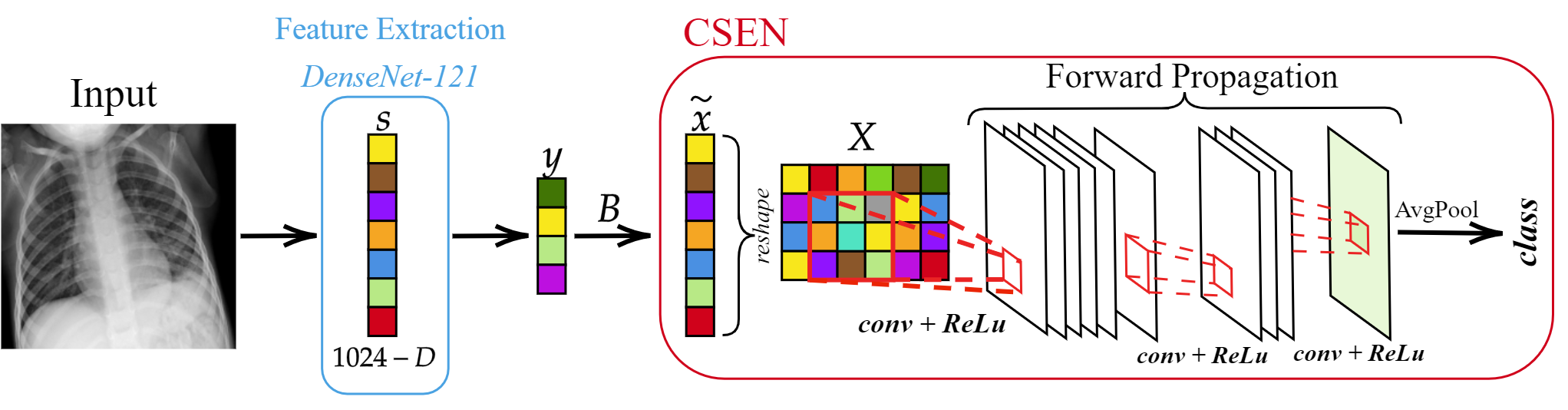}
    \caption{The CSEN approach for the early COVID-19 detection from chest X-ray images.}
    \label{fig:csen}
\end{figure*}

\subsubsection{Sparse Representation based Classification (SRC)}\label{src-section}
In SRC, when a query sample $\mathbf{y}$ is introduced, it is expected that the estimated sparse code $\hat{\mathbf{x}}$ should have a sufficient number of non-zero coefficients in the locations which correspond to samples in the dictionary, $\mathbf{D}$, with the same class of the query $\mathbf{y}$. SRC techniques are used in many different classification tasks such as face recognition in \cite{SRC2}, hyperspectral image classification \cite{hyperspecral}, and human action recognition \cite{human-action}. In the following, we briefly give more information about the SRC scheme.

Since in representation-based classification techniques, the signal $\mathbf{x}$ may not be exactly $k$-sparse due to the correlation between the samples in the dictionary, one alternative approach may be to corrupt the original scheme with an additive noise, $\mathbf{z}$, : $\mathbf{y} =\mathbf{D} \mathbf{x} + \mathbf{z}$. In this case, stable recovery of the sparse signal can still be possible even though the exact recovery is not possible, where $\hat{\mathbf{x}}$ obeys $\left \| \mathbf{x }- \hat{\mathbf{x}} \right \| \leq \kappa \left \| \mathbf{z} \right \|$ holds for the stable solution for a small $\kappa$ constant. For example, using the Lasso formulation:
\begin{equation}
\label{lasso}
    \min_\mathbf{x}  \left \{ \left \|  \mathbf{D}\mathbf{x}-\mathbf{y} \right \|_2^2 + \lambda \left \|\mathbf{ x} \right \|_1  \right \}    
\end{equation}
it is shown in \cite{lasso-stable} that it is possible to recover the partial or exact $\mathbf{x}$ in noise-free or noisy conditions, respectively.

The correlation between the samples of different classes has led many approaches to use different strategy instead of solving \eqref{lasso} directly. Accordingly, \cite{SRC2} proposes to use a four-step approach as follows:  i) The atoms of $\mathbf{D}$ and the query sample $\mathbf{y}$ are normalized to have unit $\ell_2$-norm, (ii) Perform sparse recovery: $\hat{\mathbf{x}} = \arg \min_{\mathbf{x}} \left \|\mathbf{ x} \right \|_1 \text{s.t} \left \| \mathbf{y} - \mathbf{D} \mathbf{x} \right \|_2 $, (iii) Compute the residuals using the corresponding estimated coefficient $\mathbf{\hat{x}_i}$ for the class $i$: $\mathbf{e_i} = \left \| \mathbf{y} - \mathbf{D_i}  \mathbf{\hat{x}_i} \right \|_2$ (iv) Class prediction: $\text{Class}\left ( \mathbf{y} \right ) = \arg \min \left ( \mathbf{e_i} \right )$. Since such a four-step approach brings additional improvements in the performance, many SRC studies follow a similar approach such as \cite{human-action, hyperspecral}.

\subsubsection{Collaborative Representation based Classification (CRC)}

The study in \cite{collaborative} proposes to use $\ell_2$-minimization instead of $\ell_1$-minimization in \eqref{lasso}:
\begin{equation}
    \mathbf{\hat{x} }= \arg \min_{\mathbf{x}} \left \{ \left \| \mathbf{y }- \mathbf{D}\mathbf{x }\right \|_2^2  + \lambda \left \| \mathbf{x} \right \|_2^2   \right \}
\end{equation}
 Hence, $\hat{\mathbf{x}}$ can be computed from the derived closed-form solution as $\hat{\mathbf{x}} = \left ( \mathbf{D}^T  \mathbf{D} + \lambda \mathbf{I}_{n \times n}  \right )^{-1} \mathbf{D}^T  \mathbf{ y}$. In other words, instead of searching for a sparse solution, this approach utilizes collaborative representation (as CRC term states) among the atoms of the dictionary due to the least-square sense minimization approach. Consequently, CRC is particularly faster compared to iterative $\ell_1$-minimization recovery algorithms. The CRC is used in \cite{collaborative} by modifying the second step of the previously mentioned four-step solution in Section \ref{src-section} by changing the estimation of $\hat{\mathbf{x}}$ with its closed-form estimation. They also report in \cite{collaborative} that CRC performances on different classification problems are comparable with $\ell_1$-minimization based approaches (even better than some other approaches) for high compression rates. In this work, as presented in Fig. \ref{fig:src_crc}, we use both SRC and CRC approaches to provide comparative evaluations.

\subsubsection{Convolutional Support Estimator Networks (CSENs)}

If the aim is to compute the support set rather than the exact signal recovery, then, a compact support estimator should be sufficient for this task. Moreover, in traditional approaches where signal recovery is initially performed and then $\hat{\Lambda}$ is computed, as discussed in Section \ref{preliminaries}, the performance of SE depends on the recovery performance, which is not guaranteed in noisy cases or if $\mathbf{\hat{x}}$ is not exactly sparse (e.g., for representation-based classification problems in which CRC classifier utilizes from this case as stated in \cite{collaborative}). As the recovery of partial \cite{Volkan,SE3, partial2} or complete \cite{exact4,SE1,exact1, Volkan} $\Lambda$ is still possible in these cases, it is shown in \cite{CSEN} that CSEN performs well in SE for these cases compared to traditional methods where a SR technique is first applied on $\mathbf{y}$ to compute $\mathbf{\hat{x}}$, then a thresholding is made over $\mathbf{\hat{x}}$ to estimate support set, $\hat{\Lambda}$.

The proposed CSEN network in \cite{CSEN} aims to compute direct mapping from test sample $\mathbf{y}$ to its corresponding support set $\mathbf{\hat{\Lambda}}$. Hence, the CSEN approach is faster than $\ell_1$-minimization techniques that work in an iterative manner. Moreover, CSENs have compact configurations with few convolutional layers which also contribute to computational efficiency. For example, ReconNet proposed in \cite{reconnet} originally for the signal recovery problem requires deeper network structures compared to the SE task. For the SE, another alternative is to use MLPs as the estimator networks. Similarly, such network is used in \cite{Lamp} for the recovery problem. However, for SE tasks, it is observed in \cite{CSEN} that using MLPs decreases the generalization capability and robustness to noise. Overall, thanks to their compact structures, CSENs can learn from a limited number of labelled data which is exactly the case for an advance warning system for COVID-19.

Accordingly, a SE network should compute a binary mask $\mathbf{v} \in \left \{ 0,1 \right \}^n$:
\begin{subnumcases}
{v_i =}
   1    \hfill & \text{ if $ i \in \Lambda $ } \\
   0 & \text{ else },
\end{subnumcases}
Thus, the support set would be ${\Lambda} = \left \{  i \in  \left \{  1,2,..,n\right \} : {v}_i =1   \right \}$. Correspondingly, the CSEN networks, $\mathcal{P} \left ( \mathbf{y}, \mathbf{D} \right )$, produce an output vector $\mathbf{p}$ such that $p_i \in \left [ 0,1 \right ]$ is the probability of each index being in ${\Lambda}$. Then, the estimated support set, $\hat{\Lambda} = \left \{  i \in  \left \{  1,2,..,n\right \} : p_i > \tau   \right \}$, can be computed by thresholding $\mathbf{p}$ with a fixed threshold, $\tau$.

On the other hand, the input of the CSEN is the proxy $\mathbf{\tilde{x}}$ which is a coarse estimation of $\mathbf{x}$ as $\mathbf{\tilde{x}=D^Ty}$ or $\left ( \mathbf{D}^T \mathbf{D} + \lambda \mathbf{I} \right )^{-1} \mathbf{D}^T\mathbf{y}$. Inference on the proxy of $\mathbf{x}$ is investigated by several studies for different applications, for example, studies in \cite{degerli,inference, dat} perform classification of compressively sensed images using the proxy as input of reconstruction-free frameworks.

Since CSEN networks consist of 2-D convolutional layers as illustrated in Fig. \ref{fig:csen}, the proxy $\mathbf{\tilde{x}}$ is reshaped to a 2-D plane. Then, it is convolved with the weight kernels, $\mathbf{W_1}$, connecting the input layer to the next layer with $N$ filters to form the input of the activation with the summation of weight biases $\mathbf{b_1}$:
\begin{equation}
\mathbf{F_1} = \{\text{S}(\text{ReLu}(b_1^i + \mathbf{w}_1^i * \Tilde{\mathbf{x}}))\}_{i=1}^{N},
\end{equation}
where $\text{ReLu}(x) = \text{max}(0, x)$, and $\text{S}(.)$ is the up- or down-sampling operation. Hence, the $k^{\rm {th}}$ feature map of layer $l$ can be given as,
\begin{equation}
    \mathbf{f_l^k} = \textsc{S}(\textsc{ReLu}(b_l^k + \sum_{i=1}^{N_{l-1}}\textsc{conv2D}( \mathbf{w}_l^{i,k}, \mathbf{f}_{l-1}^i, '\textsc{ZeroPad}'))).
\end{equation}
Overall, \textit{L} layer CSEN will have the trainable weight and bias parameters $\{\mathbf{w}, b\}$ as follows: $\mathbf{\Theta_{CSEN}}=\big\{ \{\mathbf{w}_1^i, b_1^i\}_{i=1}^{N_1}, \{\mathbf{w}_2^i, b_2^i\}_{i=1}^{N_2}, ... \{\mathbf{w}_L^i, b_L^i\}_{i=1}^{N_L}\big\}$.

Since samples from the same class are grouped together in the representation-based classification, a group sparsity term can be introduced in $\ell_1$-minimization problem given in \eqref{lasso} as follows,
\begin{equation}
    \min_\mathbf{x}  \left \{ \left \|  \mathbf{D}\mathbf{x}-\mathbf{y} \right \|_2^2 + \lambda \sum_{i=1}^{c}\left \|\mathbf{x_{G_i}} \right \|_2  \right \} 
\end{equation}
where the group of coefficients is represented by $\mathbf{x_{G_i}}$ for $i^{\rm {th}}$ class. Thus, the cost function would be the following for an SE network:
\begin{equation}
\label{cost}
E(\mathbf{x}) =  \sum_p (\mathcal{P}_{\Theta}\left (\mathbf{\Tilde{x}} \right )_p- v_p)^2 + \lambda \sum_{i=1}^{c}\left \|\mathcal{P}_{\Theta}\left (\mathbf{\Tilde{x}} \right )_{G_i} \right \|_2.
\end{equation}
where $v_p$ is the true binary mask indicating the sparse codes of $\mathbf{x}$ and the output of the network is $\mathcal{P}_{\Theta}\left (\mathbf{\Tilde{x}} \right )_p$ at $p^{\rm {th}}$ pixel. Although such a regularization technique may increase the classification performance since it forces the network to produce supports grouped together, the cost function in \eqref{cost} is approximated because of its computational complexity in CSEN by inserting average pooling layers after the last convolutional layer. Afterwards, the categorical cross-entropy is calculated as the cost of CSEN using the produced class probabilities obtained after SoftMax operation. Consequently, the input output pair for the training of CSEN is $(\mathbf{\Tilde{x}}, \text{class}(y))$.

Since CSEN takes reshaped proxy $\mathbf{\tilde{x}}$ as input, the corresponding indices of the atoms in the dictionary, $\mathbf{D}$, are re-ordered to make sure that samples from the same classes are grouped together in the reshaped 2-D plane. Note the fact that the grouped samples would have different sizes depending on the number of dictionary samples and classes which also determines the definite input, output mask, and average pooling stride sizes of the CSEN. The modified CSEN configurations with dictionary sizes for the two-class early detection problem will be given in Section \ref{experimental}. The overall framework of CSEN in early detection of COVID-19 is presented in Fig. \ref{fig:csen}.

\subsubsection{Multi-Layer Perceptrons (MLPs)}
\begin{figure}[b!]
    \centering
    \includegraphics[width=0.48\textwidth]{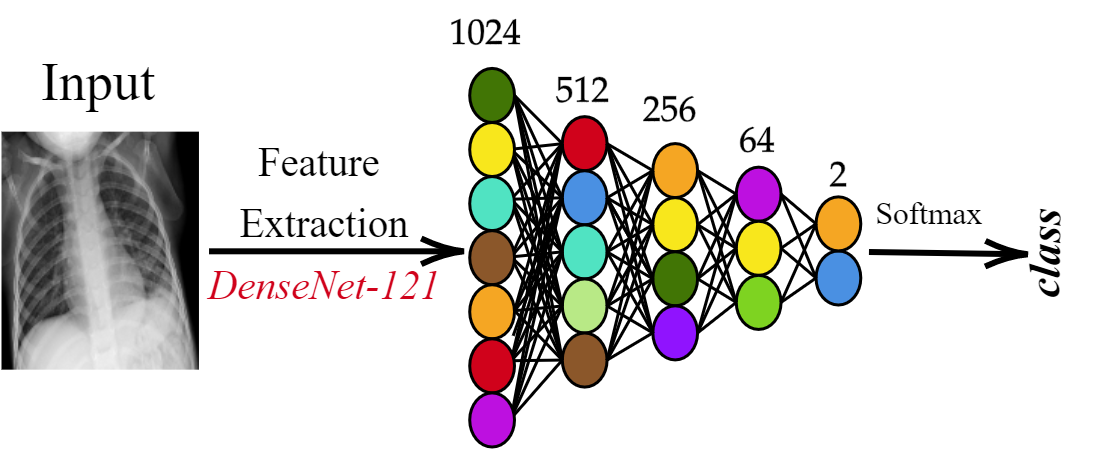}
    \caption{MLP framework used for the early detection of COVID-19.}
    \label{fig:mlp}
\end{figure}

The MLP network used for the early detection task consists of 3-hidden layers, where the details of its structure are depicted in Fig. \ref{fig:mlp}. Such network architecture is determined by testing different network topologies with shallower or deeper networks by performing parameter search. In the early detection problem with MLP based classification, we follow a slightly different approach compared to other presented compact classifiers. The neurons connecting the input layer to the first hidden layer are first initialized with $\mathbf{A} \in \mathbb{R}^{m \times d} $ which is the PCA matrix computed for dimensionality reduction task of other classifiers. Then, all layers are trained together including the PCA initialized layer. Such an approach provides a slightly improved performance and fair comparison with other classifiers used in this study since they also utilize the PCA technique.

\subsubsection{Support Vector Machine (SVM)}

The SVM topology is selected by performing grid-search to find optimal hyper-parameters such as kernel type and its parameters with the following setting and variations: the kernel function \{linear, polynomial, radial basis function (RBF)\}, kernel scale ($\gamma$ parameter for the RBF kernel) in the range $[10^{-3}, 10^{3}]$ incremented in log-scale, polynomial order \{$2$, $3$, $4$\}, and box constraint ($C$ parameter) with log-scale in the range $[10^{-3}, 10^3]$. Since this is a detection (binary classification) problem, a single SVM suffices for the task.

\subsubsection{k-Nearest Neighbor (k-NN)}

Similarly, the optimal parameters are searched in the pre-defined grid to build a k-NN classifier with the best configuration. Accordingly, in the search space, the k-values are varied with log-scale in the range $[1, N_o/2]$, where $N_o$ is the number of observations in the train set, and the evaluated $11$ different distance metrics are as follows: Euclidean, standardized Euclidean, correlation, City-block, cosine, Chebyshev, Hamming, Minkowski, Mahalanobis, Jaccard, and Spearman.

\subsection{Deep Learning based Early Detection}
Deep Learning methods, or specifically deep CNNs have achieved elegant results in many computer vision tasks. Hence, in this study, their learning capability should be investigated for the early detection of COVID-19. Contrary to the aforementioned compact classifiers, deep CNNs do not need a prior feature extraction step since CNNs combine feature extraction and classification in a single learning body and jointly optimize them. In this group, we investigate three recent deep configurations: DenseNet-121 \cite{DenseNet}, ResNet-50 \cite{resnet50}, and Inception-v3 \cite{inception}. Accordingly, DenseNet-121, ResNet-50, and Inception-v3 models are trained over the Early-QaTa-COV19 dataset starting from their ImageNet weights.

\section{Experimental Results}
\label{experimental}

In this section, first the benchmark dataset released along with this study is introduced, and then the experimental setup is presented. Finally, we provide an extensive set of comparative evaluations over early detection performances of the state-of-the-art methods covered in this study.

\subsection{Benchmark Datasets}
\subsubsection{QaTa-COV19 dataset} The researchers of Qatar University and Tampere University have compiled the largest COVID-19 dataset, called QaTa-COV19. In the dataset, there are 2951 COVID-19 positive X-ray images that are collected from various publicly available sources including BIMCV-COVID19+ \cite{vaya2020bimcv}, Hannover Medical School and Institute for Diagnostic and Interventional Radiology \cite{covidimage}, Italian Society of Medical and Interventional Radiology (SIRM) COVID-19 Database \cite{CovidDataSet2}, Chest Imaging (Spain) at thread reader \cite{CovidDataSet4}, Radiopaedia \cite{CovidDataSet3}, and news-portals and online articles. The authors have also performed the tedious task of indexing and collecting of X-ray images from various published and preprint articles from China, USA, Italy, Spain, South Korea, Taiwan for COVID-19 positive cases as well as online news-portals (until 20th of April, 2020). 

Besides COVID-19 cases, there are 12 544 normal X-ray images which are collected from various sources: Padchest dataset \cite{bustos2020padchest}, RSNA pneumonia detection challenge dataset \cite{RSNA}, Indiana Network for Patient Care \cite{indiana}, from the Department of Health and Human Services of Montgomery County and from Shenzhen Hospital released in \cite{chinaUS}, and the Kaggle chest X-ray database \cite{2018chest}.

Correspondingly, both normal and COVID-19 samples are from different gender, age-groups, ethnicity, and country. Note the fact that since collected X-ray images are from various publicly available datasets in different formats, we have removed over-exposed and low-quality samples, and duplicated samples are also removed using the structural similarity index measure (SSIM).

\subsubsection{Early-QaTa-COV19 dataset}
This dataset is formed by selecting the early stages of COVID-19 pneumonia from the enlarged version of QaTa-COV19 dataset that consists of 4603 COVID-19 positive cases. Accordingly, constructed Early-QaTa-COV19 consists of 1065 early-stage COVID-19 samples (no and very limited infection signs on 801 and 264 images, respectively) labelled by the MDs, and 12 544 samples for control (normal) class. Note the fact that the selected samples are not quantified based on their X-ray acquisition times, since they are selected from COVID-19 positive patients by visual inspection; and hence the early signs of the disease may be present at any time independent from the patients' symptoms. In fact, patients can exhibit no symptoms at all, that is, they can be asymptomatic. The dataset is highly unbalanced and this particularly makes the early detection task harder for advance warning. Furthermore, a high inter-class similarity exists in the dataset as shown in Fig. \ref{fig:interclass}. Finally, as illustrated in Fig. \ref{fig:intraclass}, there is a high intra-class dissimilarity among the images since they are compiled from different sources.

\subsection{Experimental Setup}
The comparative methods are evaluated by a $5$-fold cross-validation (CV) scheme over the Early-QaTa-COV19 dataset. We have resized chest X-ray images to $224\times224$ in order to fit the input dimensions to the state-of-the-art deep network topologies. Table \ref{foldstable} shows the number of samples in each fold, which we split the data into training and test (unseen folds) sets by $80\%$ and $20\%$, respectively. Accordingly, we also provide an initial version of Early-QaTa-COV19 as a sub-set of the original dataset. This version of the benchmark dataset is proposed specifically for the methods that are well-suited for scarce data.

\begin{table}[h!]
\centering
\caption{Number of samples per class and fold.}
\begin{subtable}{.48\textwidth}
\centering
\caption{Early-QaTa-COV19 Dataset}
\begin{tabular}{|c|c|c|c|}
\hline
\rowcolor[gray]{.95}Class & Total Samples & \begin{tabular}[c]{@{}c@{}}Training Samples\end{tabular} & \begin{tabular}[c]{@{}c@{}}Test Samples\end{tabular} \\ \hline \hline
\begin{tabular}[c]{@{}c@{}}Early Stage\\ COVID-19\end{tabular} & 1065 & 852 & 213 \\ \hline
\rowcolor[gray]{.95}Normal & 12 544 & 10 035 & 2509 \\ \hline 
\end{tabular}
\label{foldsa}
\end{subtable}

\bigskip
\noindent
\begin{subtable}{.48\textwidth}
\centering
\caption{Initial Early-QaTa-COV19 Dataset}
\begin{tabular}{|c|c|c|c|}
\hline
\rowcolor[gray]{.95}Class & Total Samples & \begin{tabular}[c]{@{}c@{}}Training Samples\end{tabular} & \begin{tabular}[c]{@{}c@{}}Test Samples\end{tabular} \\ \hline \hline
\begin{tabular}[c]{@{}c@{}}Early Stage\\ COVID-19\end{tabular} & 175 & 140 & 35 \\ \hline
\rowcolor[gray]{.95}Normal & 1579 & 1263 & 316 \\ \hline 
\end{tabular}
\label{foldsb}
\end{subtable}
\label{foldstable}
\end{table}

Since the dataset is highly unbalanced, we have balanced the training set by augmenting the data in order to have an equal number of samples in each class. However, such limited data augmentation for balancing is not enough for deep CNNs when they are trained over the initial version of the dataset. Hence, we augmented the training samples up to 20 070 X-ray images for data balancing on Early-QaTa-COV19, whereas data augmentation yields 20K images for training deep CNNs on Initial Early-QaTa-COV19. The data augmentation is performed by Image Data Generator in Keras. We have augmented the X-ray images by randomly rotating in 10 degrees of range and randomly shifting them horizontally and vertically by $10\%$. The blank sections, after rotating and shifting, are filled by the \textit{"nearest"} mode.

For the CSEN approaches used in this study, we follow the proposed configurations in \cite{CSEN}. Accordingly, there are two compact networks: CSEN1 and CSEN2. CSEN1 has only two hidden convolutional layers with 48 and 24 neurons, respectively, whereas CSEN2 consists of additional max-pooling and transposed-convolutional layers with $24$ neurons. Both networks use Rectified Linear unit (ReLu) activation functions and $3\times3$ filter sizes. In addition to this setup, the ReconNet \cite{reconnet} approach is modified to perform the SE task as a deep version of the CSEN framework. ReconNet is originally proposed for the signal recovery problem as a non-iterative alternative to the traditional approaches, and it achieves the state-of-the-art performance levels in compressive sensing applications as shown in \cite{reconnet}. The modified ReconNet for SE has 6 fully convolutional layers, and it does not have a denoiser layer as the first block and Block-matching and 3D filtering (BM3D) operation (see \cite{reconnet}) at the output. The input-output pair is also different ($\mathbf{\Tilde{x}}, \text{class}(\mathbf{y})$) to train the network as the CSEN type of approach for the early detection problem. Accordingly, its last layer is modified by inserting an average-pooling layer to mimic the cost in \eqref{cost}, and SoftMax to produce class probabilities.

The experimental evaluations of SRC, CRC, and k-NN are performed on a PC with Intel ® i$7-8650$U CPU and 32 GB system memory with MATLAB version 2019a, whereas SVM is implemented with the same computer setup but in Python. In the regularized least square solution of CRC, the regularization parameter is searched in the range $[10^{-13}, 10^{3}]$ incremented in log-scale, then, it is fine-tuned with $\lambda^{*}/10$ steps. For the hyper-parameter selection of k-NN and SVM classifiers, grid search is performed using another $5$-fold stratified CV over the training sets of the previously explained CV folds. Other approaches: MLP, CSEN, and deep learning methods are implemented with the Tensorflow library \cite{abadi2016tensorflow} using Python on NVidia ® TITAN-X GPU card. The training procedures of MLP, CSEN and deep CNNs are performed using ADAM optimizer \cite{adam} with their proposed default momentum update parameters as $\beta_1 = 0.9$ and $\beta_2 = 0.999$ using categorical cross-entropy loss function. CSEN is trained for only $15$ Back-Propagation epochs with a learning rate, $\alpha=10^{-4}$, and a batch size, $32$. On the other hand, the MLP network and deep learners are trained with $\alpha = 10^{-5}$ both for $10$ epochs and with batch size $32$.

\subsection{Results}

For a possible advance warning system that can detect COVID-19 in the early stages, we have analyzed and evaluated several ML approaches including compact classifiers: SRC, CRC, CSEN, MLP, SVM, and k-NN, and deep CNNs: DenseNet-121, ResNet-50, and Inception-v3 networks. For SRC approach, we have investigated 8 different solvers: OMP \cite{fast}, Dalm \cite{fast}, L1LS \cite{l1ls}, ADMM \cite{ADMM}, Homotopy \cite{homotopy}, GPSR \cite{gpsr}, Palm \cite{fast}, and $\ell_1$-magic \cite{l1magic}. In this study, we report SRC results from only Dalm and Homotopy solvers since others show poor performance in the detection task (provides $< 80\%$ sensitivity).

In the representation-based classification approaches, the dictionary is constructed by using all samples from the balanced training set of each fold. Hence, $\mathbf{\Phi}$, has 10 035 and 1263 samples from each class of Early-QaTa-COV19 and its initial version, respectively. Next, the PCA matrix, $\mathbf{A}$, is applied for the dimensionality reduction, where the compression ratio is $\text{CR}=m/d = 0.5$. The equivalent dictionary, $\mathbf{D}$, would have the size of $512 \times 20 070$ and $512 \times 2526$ for the initial version. On the other hand, since CSEN networks need additional training samples, we choose only $625$ samples per class to construct $\mathbf{\Phi}$ and use the remaining samples from the training set of each fold to train CSENs in the experiments. Consequently, the corresponding denoiser matrix $\mathbf{B} = \left ( \mathbf{D}^T \mathbf{D} + \lambda \mathbf{I} \right )^{-1} \mathbf{D}^T$ of size $1250 \times 512$ is used to perform coarse estimation for CSEN. The resulted $\mathbf{\Tilde{x}} = \mathbf{By}$, $\mathbf{\Tilde{x}} \in \mathbb{R}^{n=1250}$, is reshaped to 2-D plane with the size of $25 \times 50$ in such a way that support sets from the corresponding classes are grouped together which is then fed to CSENs.

\subsubsection{Results on Early-QaTa-COV19}

The early detection performances of the compact classifiers are given with their $95\%$ confidence intervals (CIs) in Table \ref{results_compact}. Accordingly, CI can be estimated for each performance metric as follows: $r = z \sqrt{\text{accuracy}(1 - \text{accuracy}) / N}$, where $N$ is the number of samples for that particular performance metric, and $z$ is the level of significance that is 1.96 for $95\%$ CI. Consequently, it is expected that the CIs for sensitivity measure are larger due to the unbalanced data. The presented results clearly indicate that CSENs achieve the top sensitivities among other compact classifiers with acceptable specificity rates. In particular, CSEN2 configuration achieves over $97\%$ sensitivity with a high specificity ($>95\%$). The same denoiser matrix is also used in the CRC method which is reported separately as the CRC-light version to observe if the CSEN approach brings performance improvement. Since representation-based classification approaches are known to perform well in the limited-data scenarios, the second competitor is indeed the CRC approach from the first group as expected. Note the fact that the CRC method provides comparable classification performance for the early detection with SRC methods. This may be explained in the following way; in representation-based classification problems, it may not be the sparsity that brings the information about the class but the collaborative representation among the samples in the dictionary. Similar findings are reported in \cite{collaborative} for the face recognition problem.

\begin{table}[hb!]
\centering
\caption{The average performances and their $95\%$ confidence intervals (CIs) of different approaches over 5-folds on Early-QaTa-COV19 for the early detection of COVID-19 pneumonia from the normal chest X-ray images. CRC-light uses the same $\mathbf{\Phi}$ with CSENs, whereas CRC uses the complete training set.}
\small
\begin{subtable}{.48\textwidth}
\caption{Average detection performances of compact classifiers.}
\centering
 \resizebox{0.95\textwidth}{!}{
\begin{tabular}{|c|c|c|c|}
\hline
\rowcolor[gray]{.85} Method & Accuracy & Sensitivity & Specificity\\ \hline \hline
SRC-Dalm & 0.9852 \textpm\ 0.002 & 0.8864 \textpm\ 0.019 & 0.9935 \textpm\ 0.001 \\ \hline
\rowcolor[gray]{.95} SRC-Hom. & 0.9778 \textpm\ 0.025 & 0.9211 \textpm\ 0.016 & 0.9826 \textpm\ 0.002 \\ \hline
CRC-light & 0.9730 \textpm\ 0.003 & 0.9559 \textpm\ 0.012 & 0.9744 \textpm\ 0.003 \\ \hline
\rowcolor[gray]{.95} CRC & 0.9678 \textpm\ 0.003 & 0.9155 \textpm\ 0.017 & 0.9723 \textpm\ 0.003 \\ \hline
CSEN1 & 0.9513 \textpm\ 0.004 & 0.970 \textpm\ 0.01 & 0.9497 \textpm\ 0.004 \\ \hline
\rowcolor[gray]{.95}CSEN2 & 0.9566 \textpm\ 0.003 & 0.9728 \textpm\ 0.010 & 0.9552 \textpm\ 0.004 \\ \hline
ReconNet & 0.9322 \textpm\ 0.004 & 0.9662 \textpm\ 0.011 & 0.9293 \textpm\ 0.005 \\ \hline
\rowcolor[gray]{.95}MLP & 0.9699 \textpm\ 0.003 & 0.9352 \textpm\ 0.015 &	0.9728 \textpm\ 0.003 \\ \hline
SVM & 0.9830 \textpm\ 0.002 & 0.8892 \textpm\ 0.019 & 0.9910 \textpm\ 0.002 \\ \hline
\rowcolor[gray]{.95}k-NN & 0.9741 \textpm\ 0.003 & 0.9305 \textpm\ 0.015 & 0.9778 \textpm\ 0.003 \\ \hline
\end{tabular}}
\label{results_compact}
\end{subtable}

\bigskip
\noindent
\begin{subtable}{.48\textwidth}
\caption{Average detection performances of deep CNNs.}
\centering
\resizebox{0.95\textwidth}{!}{
\begin{tabular}{|c|c|c|c|}
\hline
\rowcolor[gray]{.85} Method & Accuracy & Sensitivity & Specificity\\ \hline \hline
DenseNet-121 & 0.9937 \textpm\ 0.001 & 0.9502 \textpm\ 0.013 & 0.9974 \textpm\ 0.001 \\ \hline
\rowcolor[gray]{.95} Inception-v3 & 0.9791 \textpm\ 0.002 & 0.8469 \textpm\ 0.022 & 0.9904 \textpm\ 0.002 \\ \hline
ResNet-50 & 0.9884 \textpm\ 0.002 & 0.9155 \textpm\ 0.017 & 0.9946 \textpm\ 0.001 \\ \hline 
\end{tabular}}
\label{results_deep}
\end{subtable}
\label{results_approaches}
\end{table}

\begin{table}[h!]
\centering
\caption{Leading \textit{compact} CSEN and \textit{deep} DenseNet-121 models' cumulative confusion matrices on Early-QaTa-COV19.}
\begin{subtable}{.48\textwidth}
\centering
\caption{CSEN2 Confusion Matrix}
\small
    \begin{tabular}{|c|c|c|c|}
\hline
\multicolumn{2}{|c|}{\multirow{2}{*}{\textbf{CSEN2}}} & \multicolumn{2}{c|}{Predicted} \\ \cline{3-4} 
\multicolumn{2}{|c|}{} & \multicolumn{1}{c|}{Normal} & \multicolumn{1}{c|}{COVID-19} \\ \hline
\multirow{2}{*}{\begin{tabular}[c]{@{}c@{}}Ground\\ Truth\end{tabular}} & Normal & 11 982 & 562 \\ \cline{2-4} 
 & COVID-19 & 29 & 1036 \\ \hline
\end{tabular}
\label{CMa}
\end{subtable}

\bigskip
\noindent
\begin{subtable}{.48\textwidth}
\centering
\caption{DenseNet-121 Confusion Matrix}
\small
    \begin{tabular}{|c|c|c|c|}
\hline
\multicolumn{2}{|c|}{\multirow{2}{*}{\textbf{DenseNet-121}}} & \multicolumn{2}{c|}{Predicted} \\ \cline{3-4} 
\multicolumn{2}{|c|}{} & \multicolumn{1}{c|}{Normal} & \multicolumn{1}{c|}{COVID-19} \\ \hline
\multirow{2}{*}{\begin{tabular}[c]{@{}c@{}}Ground\\ Truth\end{tabular}} & Normal & 12 511 & 33 \\ \cline{2-4} 
 & COVID-19 & 53 & 1012 \\ \hline
\end{tabular}
\label{CMb}
\end{subtable}
\label{CMs}
\end{table}

\begin{figure*}[ht!]
\centering
\includegraphics[width=1\linewidth]{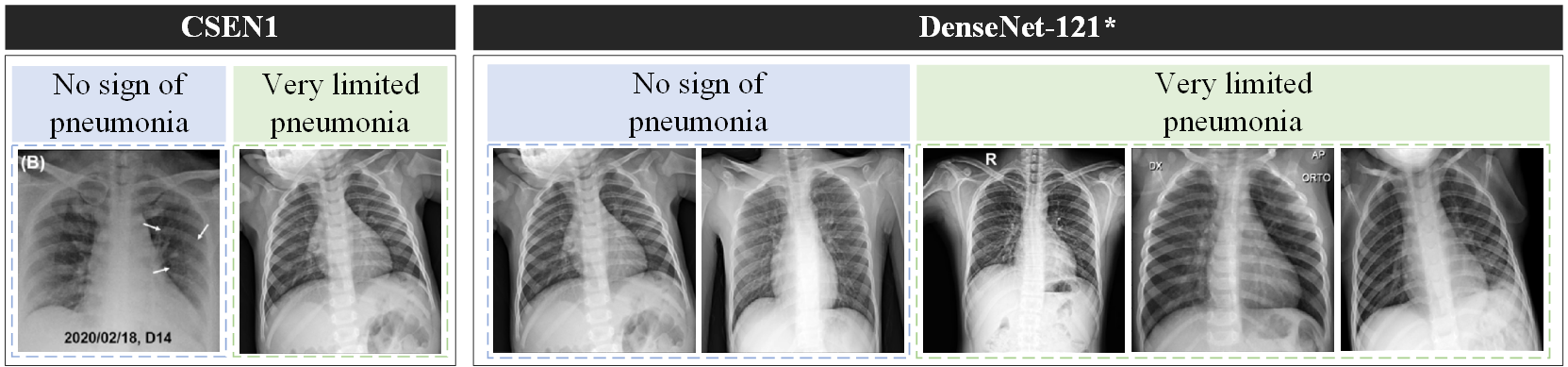}
\caption{False negatives of the CSEN1 and DenseNet-121\textsuperscript{*} which is initialized with ChestX-ray14 weights and fine-tuned on Initial Early-QaTa-COV19.}
\label{fig:FN}
\end{figure*}

Table \ref{results_deep} presents the average detection performances of deep CNNs. Although the best sensitivity is obtained using CSENs, deep CNNs can also achieve quite high sensitivity levels with almost no false alarms (specificity is $\sim1$). For example, DenseNet-121 outperforms other deep networks and produces a superior specificity of $99.74\%$ but with a reduced sensitivity than CSEN2. Moreover, the confusion matrices, cumulated over the confusion matrix of each fold's test set, are presented in Table \ref{CMs} for the two top-performers. Accordingly, CSEN2 is able to detect 24 more early COVID-19 cases than DenseNet-121.

\subsubsection{Results on Initial Early-QaTa-COV19}

Initial version of Early-QaTa-COV19 enables to benchmark methods for the cases when the data is limited. To overcome this limitation, a transfer learning technique can still be applied by the following: (i) using a suitable pre-trained model without fine tuning to extract features which will be used by a compact classifier that does not require a large amount of training data, and (ii) using the pre-trained deep model as the initialization, and then, performing fine-tuning with limited data. In this manner, in the experiments with Initial Early-QaTa-COV19, we use DenseNet-121 for extracting features of the compact classifiers with the weights learned over ChestX-ray14 dataset.

Experimental evaluations on Initial Early-QaTa-COV19 is important for a fair comparison between CSEN type of networks with CRC and SRC methods. Because, even though the dictionaries of the representation based approaches are formed by using the complete training set, the more fair comparison against CSENs can be performed by using a small training set (for all methods) but still with the same size with the dictionaries of CRC and SRC methods.

The early detection performances of the methods are given in Table \ref{results_approaches_tiny}. It is clear that CSEN1 and ReconNet can achieve the top sensitivity among all methods. However, ReconNet suffers from relatively low specificity, which indicates a substantial amount of false positives. Note the fact that the CRC method outperforms SRC methods on Initial Early-QaTa-COV19. Since the data is limited; and hence, the number of atoms in the dictionary is smaller, we have observed much higher performance gap between CRC and SRC compared to the previous experiments on Early-QaTa-COV19 and other classification problems reported in the previous studies \cite{collaborative,CSEN}. Considering the second group of approaches based on deep CNNs, DenseNet-121 is additionally initialized with ChestX-ray14 weights, and it is fine-tuned over Initial Early-QaTa-COV19. In this way, we also aim to investigate if the ChestX-ray14 dataset improves the performance of the early detection of COVID-19 by comparing the DenseNet-121 version which is trained directly from its ImageNet weights. Indeed, initialization with ChestX-ray14 weights has slightly improved the sensitivity but with a specificity of around $99.5\%$.

\begin{table}[h!]
\centering
\caption{The average performances and their $95\%$ confidence intervals (CIs) of different approaches over 5-folds on Initial Early-QaTa-COV19 for the early detection of COVID-19 pneumonia from the normal chest X-ray images. CRC-light uses the same $\mathbf{\Phi}$ with CSENs, whereas CRC uses the complete training set.}
\small
\begin{subtable}{.48\textwidth}
\caption{Average detection performances of compact classifiers.}
\centering
 \resizebox{0.95\textwidth}{!}{
\begin{tabular}{|c|c|c|c|}
\hline
\rowcolor[gray]{.85} Method & Accuracy & Sensitivity & Specificity\\ \hline \hline
SRC-Dalm & 0.9818 \textpm\ 0.006 & 0.9371 \textpm\ 0.036 & 0.9867 \textpm\ 0.006 \\ \hline
\rowcolor[gray]{.95} SRC-Hom. & 0.9481 \textpm\ 0.010 & 0.8171 \textpm\ 0.057 & 0.9626 \textpm\ 0.009 \\ \hline
CRC-light & 0.9783 \textpm\ 0.007 & 0.9486 \textpm\ 0.033 & 0.9816 \textpm\ 0.007 \\ \hline
\rowcolor[gray]{.95} CRC & 0.9823 \textpm\ 0.006 & 0.9657 \textpm\ 0.027 & 0.9842 \textpm\ 0.006 \\ \hline
CSEN1 & 0.9635 \textpm\ 0.009 & 0.9886 \textpm\ 0.016 & 0.9607 \textpm\ 0.010 \\ \hline
\rowcolor[gray]{.95}CSEN2 & 0.9248 \textpm\ 0.012 & 0.9943 \textpm\ 0.011 & 0.9171 \textpm\ 0.014 \\ \hline
ReconNet & 0.9424 \textpm\ 0.011 & 0.9943 \textpm\ 0.011 & 0.9367 \textpm\ 0.012 \\ \hline
\rowcolor[gray]{.95}MLP & 0.9584 \textpm\ 0.009 & 0.9371 \textpm\ 0.036 &	0.9607 \textpm\ 0.010 \\ \hline
SVM & 0.9681 \textpm\ 0.008 & 0.9657 \textpm\ 0.027 & 0.9683 \textpm\ 0.009 \\ \hline
\rowcolor[gray]{.95}k-NN & 0.9458 \textpm\ 0.011 & 0.9257 \textpm\ 0.039 & 0.9481 \textpm\ 0.011 \\ \hline
\end{tabular}}
\label{results_compact_tiny}
\end{subtable}

\bigskip
\noindent
\begin{subtable}{.48\textwidth}
\caption{Average detection performances of deep CNNs. DenseNet-121\textsuperscript{*} is initialized with ChestX-ray14 weights.}
\centering
\resizebox{0.95\textwidth}{!}{
\begin{tabular}{|c|c|c|c|}
\hline
\rowcolor[gray]{.85} Method & Accuracy & Sensitivity & Specificity\\ \hline \hline
DenseNet-121\textsuperscript{*} & 0.9926 \textpm\ 0.004 & 0.9714 \textpm\ 0.025 & 0.9949 \textpm\ 0.004 \\ \hline
\rowcolor[gray]{.95} DenseNet-121 & 0.9949 \textpm\ 0.003 & 0.9543 \textpm\ 0.031 & 0.9994 \textpm\ 0.001 \\ \hline
Inception-v3 & 0.9937 \textpm\ 0.004 & 0.9543 \textpm\ 0.031 & 0.9981 \textpm\ 0.002 \\ \hline
\rowcolor[gray]{.95} ResNet-50 & 0.9943 \textpm\ 0.004 & 0.9600 \textpm\ 0.029 & 0.9981 \textpm\ 0.002 \\ \hline 
\end{tabular}}
\label{results_deep_tiny}
\end{subtable}
\label{results_approaches_tiny}
\end{table}

Moreover, the false-negative X-ray images are given in Fig. \ref{fig:FN}. Accordingly, DenseNet-121 initiliazed with ChestX-ray14 weights misses three more early case of COVID-19 than CSEN1, but it is able to provide much higher specificity (the sensitivity for normal X-ray images). It is observed that CSEN1 and DenseNet-121 are able to detect $39$ and $38$, respectively, out of $40$ early cases of COVID-19 that show no visible sign of COVID-19 by a human eye.

\subsubsection{Computational Complexity Analysis}

\begin{figure}[ht!]
  \centering
  \includegraphics[width=1\linewidth]{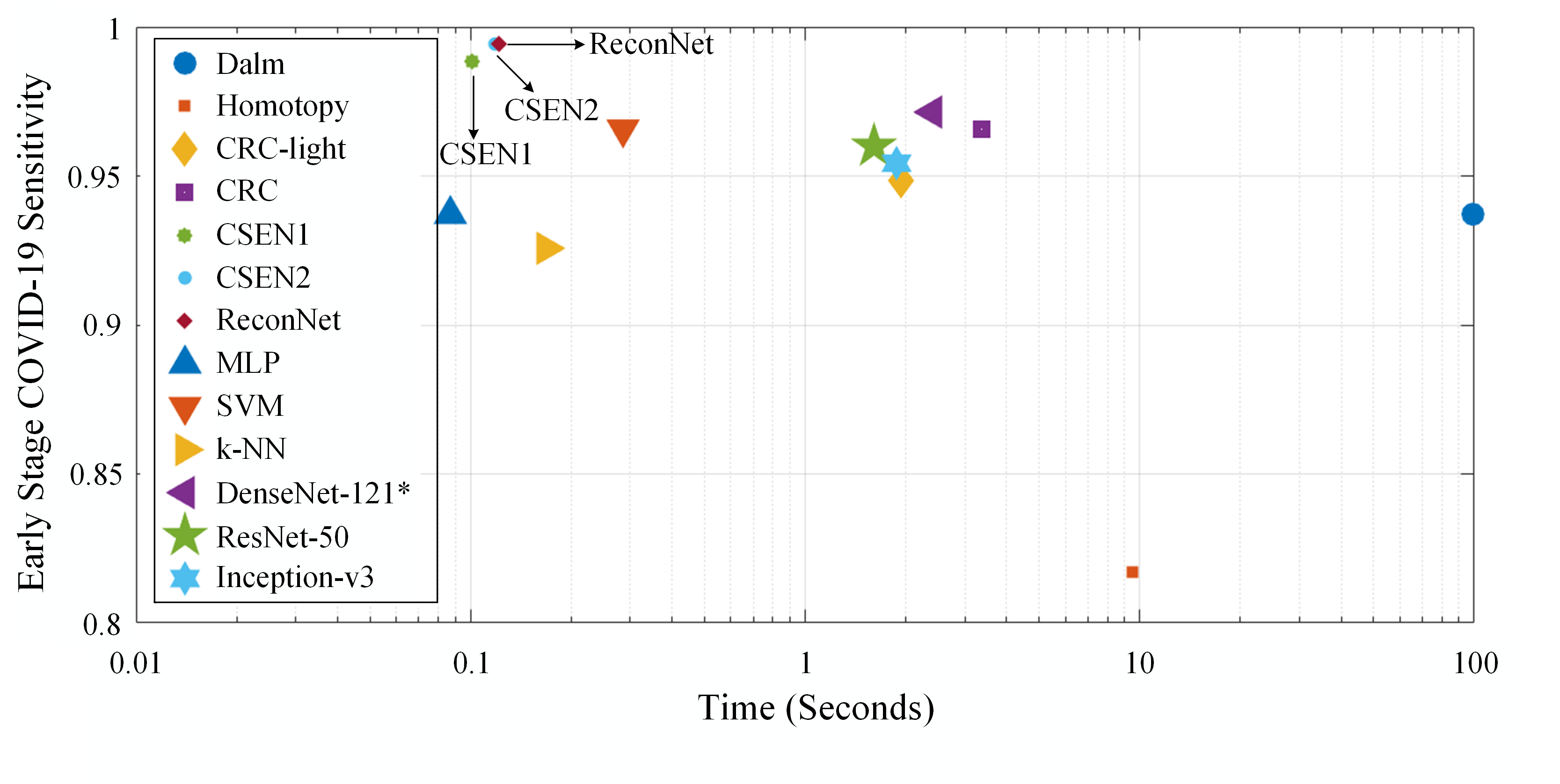}  
  \caption{Time complexity versus the sensitivity of all the evaluated classifiers. Computational times are plotted in log-scale and measured for the evaluation of test sets by averaging over $5$-folds.}
  \label{time_vs_sensitivity}
\end{figure}

To assess the computational complexity analysis of the compared methods, we first start with the number of trainable network parameters as presented in Table \ref{paramaters}. Obviously CSENs have a crucial advantage especially compared to the deep CNNs in terms of computational complexity. As for time complexity, Fig. \ref{time_vs_sensitivity} shows sensitivity versus computational time for each method evaluated in this study. Note that the computational times are evaluated on Initial Early-QaTa-COV19 dataset for representation based classification approaches including CRC, Dalm, and Homotopy since their computational times are only comparable if $\mathbf{ \Phi}$ consists of small number of atoms. Hence, the sensitivity levels on the initial version of the dataset are given in Fig. \ref{time_vs_sensitivity}. It is observed that CSEN type of networks achieve computational efficiency, and DenseNet-121 can achieve slightly inferior sensitivity and it is still faster than some of the compact classifiers.

\begin{table}[h!]
\centering
\caption{Number of trainable parameters of comparative models.}
\small
\begin{tabular}{|c|c|}
\hline
\rowcolor[gray]{.85} Model & Number of Parameters \\ \hline \hline
CSEN1 & 11,089 \\ \hline
\rowcolor[gray]{.98} CSEN2 & 16,297 \\ \hline
ReconNet & 22,914 \\ \hline
\rowcolor[gray]{.98}MLP & 672,706 \\ \hline
DenseNet-121 & 6,955,906  \\ \hline
\rowcolor[gray]{.98}ResNet-50 & 23,538,690 \\ \hline
Inception-v3 & 21,772,450 \\ \hline
\end{tabular}
\label{paramaters}
\end{table}

Note the fact that even though CRC tends to provide closed-form solution, because of the four-step classification framework, which involves residual finding as discussed in Section \ref{methods}, representation-based classification techniques suffer from the highest time complexity in general. However, CSENs use only the denoiser multiplication part of CRC, $\mathbf{\Tilde{x}} = \mathbf{By}$, which requires an insignificant time (i.e., only $~3.86$ ms for a test set averaged over $5$-folds). Finally, such inference times for deep networks are valid if they can utilize the recent GPU cards, whereas other methods do not require additional hardware and can run on an ordinary computer. This is a crucial advantage for those light-weight mobile applications with a real-time analysis requirement.

\section{Limitations of the Study}

In this study, for the first time in the literature ML approaches are applied for the early detection of COVID-19. On the other hand, it can be said that the number of COVID-19 images may still be considered as a limitation considering hundreds of thousands of samples in the existing datasets for other applications. Moreover, another important drawback: the decision of the methods is not easily accessible or interpretable at all since they are considered as black-box approaches. Of course, there are preliminary works of exploring activation maps showing which locations actually activate the networks' decision. However, there is further study needed to investigate their correctness. This will be the topic of the future work. Nonetheless, automated methods can still be employed due to their cost-effectiveness and speeds as precautionary measure while waiting other diagnostic tools such as CT-scans and RT-PCR. For example, one may utilize evaluated methods for the early detection task especially in the heavily effected places where the diagnosis based on CT-scan or RT-PCR takes time.

\section{Conclusion}
\label{conclusion}

Since there is no known specific treatment for COVID-19, the early detection of the disease plays a vital role in preventing the spreading of the pandemic. Currently, RT-PCR is widely used in the world for the diagnosis of COVID-19. However, RT-PCR tests can easily miss a positive case (false negatives) depending on the sample collection or the disease stage of the patient. As an alternative, chest CT-scans have provided satisfactory results and outperformed sensitivity levels of RT-PCR. Nevertheless, in many areas where hospitals are congested because of the pandemic, it may not be easy to access such expensive and time-consuming equipment.

X-ray acquisition, however, is cheaper, easily accessible, and the acquisition time is shorter than CT. Furthermore, X-ray imaging can be applied with a greater ease since the equipment is portable. This justifies our motivation to investigate the feasibility of an accurate, robust and fully-automatic advance warning method for COVID-19 from chest X-ray images. For this purpose, we first compiled the Early-QaTa-COV19 dataset which encapsulates the largest number of COVID-19 patients who are in the early stages. Our findings have clearly demonstrated the fact that early detection of COVID-19 infestation from X-ray images can be performed with a very high sensitivity and specificity. In other words, even though it is a difficult or sometimes impossible task for experts to detect COVID-19 infestation due to the early stage of the disease, with a proper setup and training, some particular compact and deep classifiers can accurately detect the disease with tolerable false-positives. In particular, it is observed that CSEN type of models provide the highest sensitivity levels with $>97\%$ while DenseNet-121 provides a decreased sensitivity with a higher specificity. Among those $40$ X-ray images in Initial Early-QaTa-COV19 where MDs have found no trace of COVID-19 infestation (and hence naturally would mis-diagnose all of them as "normal"), CSEN1 and DenseNet-121 can accurately identify $39$ and $38$ of them, respectively. Finally, both CSEN models have the utmost computational efficiency especially when compared to the CRC and deep networks. This makes them a feasible solution for those low-cost/low-power portable applications.

\bibliographystyle{IEEEtranDOI}
\bibliography{IEEEtran}

\begin{IEEEbiography}[{\includegraphics[width=1in,height=1.25in,clip,keepaspectratio]{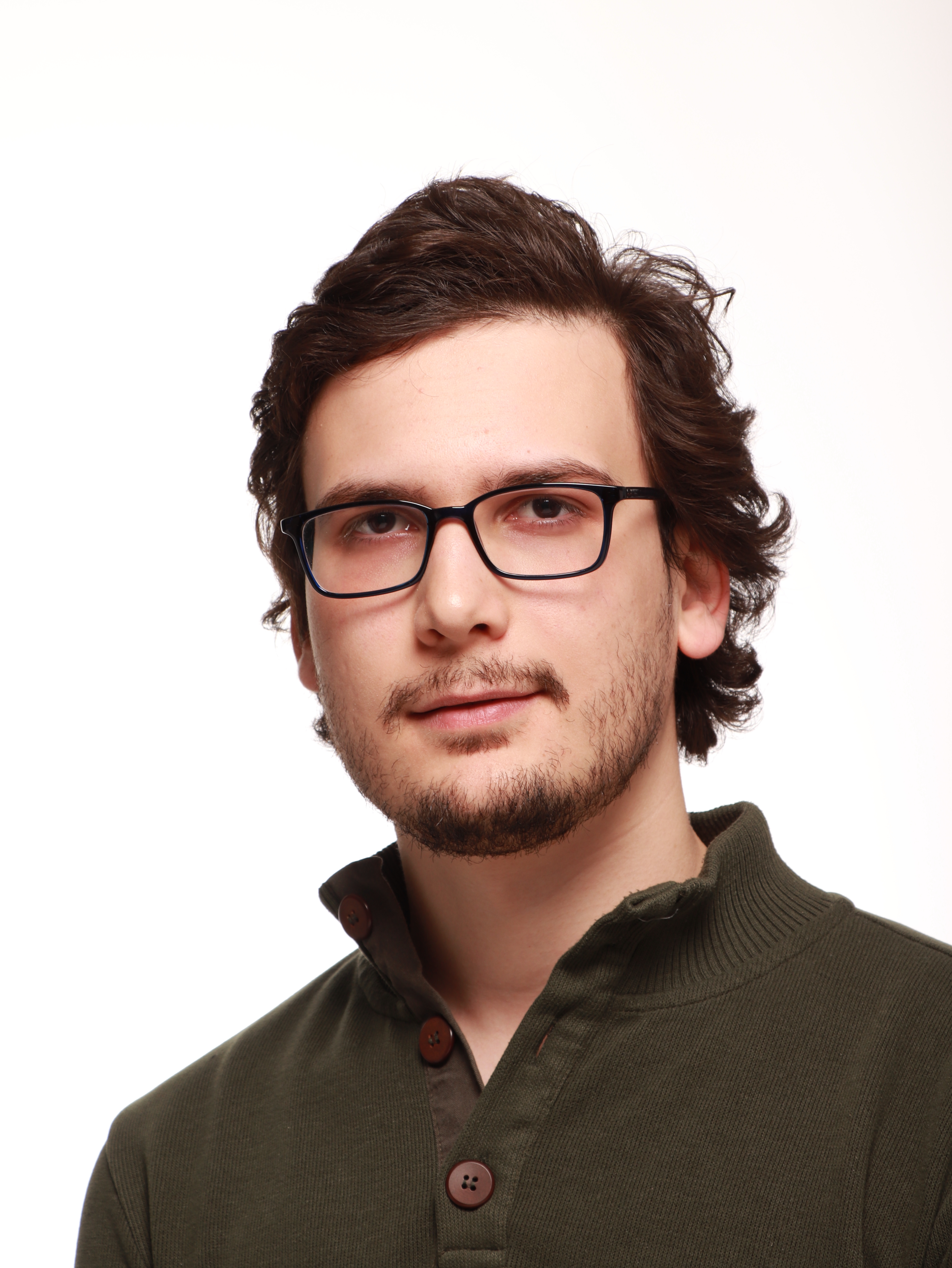}}]{Mete Ahishali} received the B.Sc. degree (Hons.) in Electrical and Electronics Engineering from Izmir University of Economics, Izmir, Turkey, in 2017, and the M.Sc. degree in Data Engineering and Machine Learning from Tampere University, Tampere, Finland, in 2019. Since 2017, he has been working as a researcher in Signal Analysis and Machine Intelligence research group under the supervision of Prof. Gabbouj, and he is currently pursing the Ph.D. degree in Computing and Electrical Engineering at Tampere University. His research interests are pattern recognition, machine learning, and semantic segmentation with applications in computer vision, remote sensing, and biomedical images.
\end{IEEEbiography}
\begin{IEEEbiography}[{\includegraphics[width=1in,height=1.25in,clip,keepaspectratio]{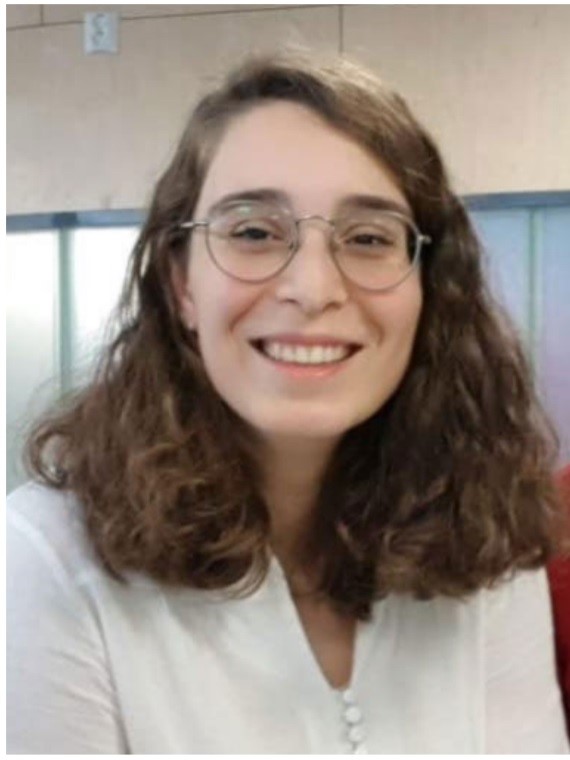}}]{Aysen Degerli} received the B.Sc. degree (Hons.) in Electrical and Electronics Engineering from the Izmir University of Economics, Izmir, Turkey, in 2017, and the M.Sc. degree (Hons.) in Data Engineering and Machine Learning from Tampere University, Tampere, Finland, in 2019, where she is currently pursuing the Ph.D. degree in Computing and Electrical Engineering. She is currently a Doctoral Researcher in Signal Analysis and Machine Intelligence research group led by Prof. M. Gabbouj with Tampere University. Her research interests include machine learning, compressive sensing, and biomedical image processing.
\end{IEEEbiography}
\begin{IEEEbiography}[{\includegraphics[width=1in,height=1.25in,clip,keepaspectratio]{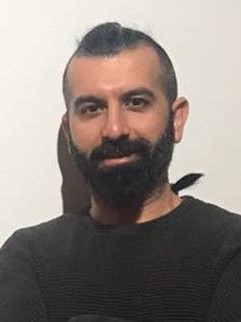}}]{Mehmet Yamac} received his B.S. degree in Electrical and Electronics Engineering from Anadolu University, Eskisehir, Turkey, in 2009 the M.S. degree in Electrical and Electronics Engineering from Bogazici University, Istanbul, Turkey, in 2014. He was research and teaching assistant at Bogazici University during 2012-2017. He is currently a Ph.D. candidate at the Department of Computing Sciences, Tampere University, Tampere, Finland. His research interests are computer and machine vision, machine learning and compressive sensing.
\end{IEEEbiography}
\begin{IEEEbiography}[{\includegraphics[width=1in,height=1.25in,clip,keepaspectratio]{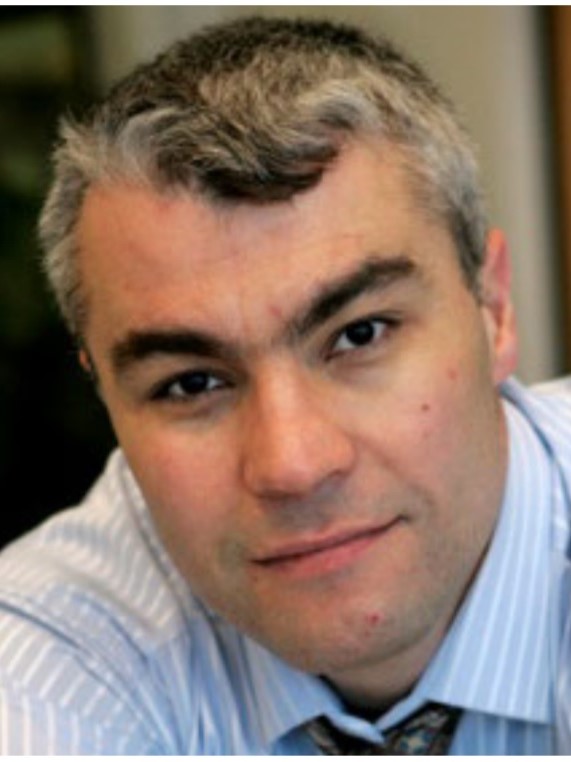}}]{Serkan Kiranyaz} (SM’13) is a Professor in Qatar University, Doha, Qatar. He published 2 books, 5 book chapters, more than 70 journal papers in high impact journals, and 100 papers in international conferences. He made contributions on evolutionary optimization, machine learning, bio-signal analysis, computer vision with applications to recognition, classification, and signal processing. Prof. Kiranyaz has co-authored the papers which have nominated or received the “Best Paper Award” in ICIP 2013, ICPR 2014, ICIP 2015 and IEEE TSP 2018. He had the most-popular articles in the years 2010 and 2016, and most-cited article in 2018 in IEEE Transactions on Biomedical Engineering. During 2010-2015 he authored the 4th most-cited article of the Neural Networks journal. His research team has won the 2nd and 1st places in PhysioNet Grand Challenges 2016 and 2017, among 48 and 75 international teams, respectively. In 2019, he won the “Research Excellence Award” and “Merit Award” of Qatar University. His theoretical contributions to advance the current state of the art in modelling and representation, targeting high long-term impact, while algorithmic, system level design and implementation issues target medium and long-term challenges for the next five to ten years. He in particular aims at investigating scientific questions and inventing cutting edge solutions in “personalized biomedicine” which is in one of the most dynamic areas where science combines with technology to produce efficient signal and information processing systems meeting the high expectation of the users
\end{IEEEbiography}
\begin{IEEEbiography}[{\includegraphics[width=1in,height=1.25in,clip,keepaspectratio]{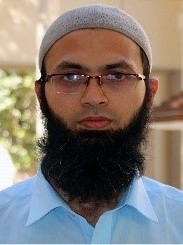}}]{Muhammad E. H. Chowdhury} received his B.Sc. degree and M.Sc. degree with record marks from the department of Electrical and Electronics Engineering at the University of Dhaka, Bangladesh. He acquired his Ph.D. degree from the University of Nottingham, U.K. at 2014. He is currently working as a full-time faculty at Electrical Engineering department, Qatar University. He worked as Post-doctoral research fellow and Hermes Fellow at the Sir Peter Mansfield Imaging Centre in the University of Nottingham, U.K. Prior to joining Qatar University, he worked in several universities of Bangladesh. His current research interests include biomedical instrumentation, machine learning, signal processing, wearable sensors, medical image analysis, computer vision, embedded system design, biomedical implants and simultaneous EEG/fMRI. He has a patent and published 48 peer reviewed journal articles, 29 conference papers and two book chapters. Dr. Chowdhury is currently running several NPRP and UREP grants from QNRF and internal grants from Qatar University along with academic and government projects. He has been involved in EPSRC, ISIF and EPSRC-ACC grants along with different national and international projects. He has worked as consultant for the projects entitled, “Driver Distraction Management Using Sensor Data Cloud (2013-14, Information Society Innovation Fund (ISIF) Asia)”. He has received ISIF Asia Community Choice Award 2013 for project entitled “Design and Development of Precision Agriculture Information System for Bangladesh”. He has recently won COVID-19 dataset award for his contribution to fight against COVID-19. He is an active member of IEEE, British Radiology, Institute of Physics, ISMRM and HBM. He is severing as an Associate Editor for IEEE Access and Review Editor for Frontiers in Neuroscience.
\end{IEEEbiography}
\begin{IEEEbiography}[{\includegraphics[width=1in,height=1.25in,clip,keepaspectratio]{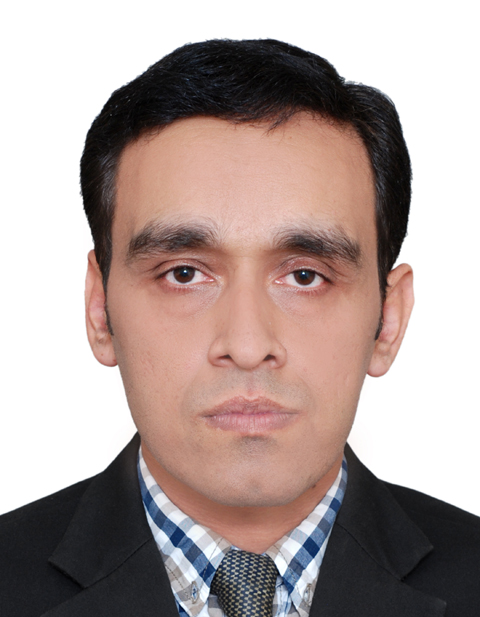}}]{Khalid Hameed} is young enthusiastic doctor with particular interest in Artificial intelligence-based work in the field of radiology. His recent publication includes X-ray imagining in COVID-19. He is affiliated with Hamad Medical corporation-Qatar and have worked at Reem Medical Center-Qatar.
\end{IEEEbiography}
\begin{IEEEbiography}[{\includegraphics[width=1in,height=1.25in,clip,keepaspectratio]{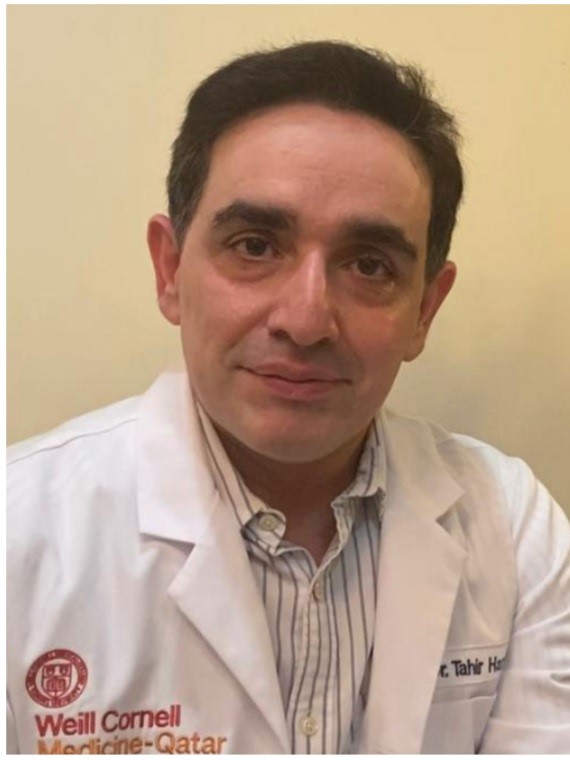}}]{Tahir Hamid} (MRCP-UK, FESC) has been working as interventional cardiology consultant at Heart Hospital, Hamad Medical Corporation, Doha Qatar. He has been trained in United Kingdom and at University of Toronto, Canada. He has about 27 publications and presented at various cardiology conferences. Currently, conducting randomized control trial about “pre-procedural fasting in patients undergoing coronary interventions” at Toronto General Hospital, University of Toronto. He has published the following study related to ECG monitoring in patients with recurrent blackouts: “Prolonged implantable electrocardiographic monitoring indicates a high rate of misdiagnosis of epilepsy - REVISE study”, Europace. Nov. 2012;14(11):1653-60. This study revolutionized the assessment of patient blackout who were wrongly labeled as epileptics and were later found after prolonged monitoring, to have cardiac rhythm related issues leading to blackouts. He is currently actively involved in the research projects at Hamad Medical Corporation. He is currently working on his project about external cardiopulmonary resuscitation devices for patients who sustain cardiac arrests.
\end{IEEEbiography}
\begin{IEEEbiography}[{\includegraphics[width=1in,height=1.25in,clip,keepaspectratio]{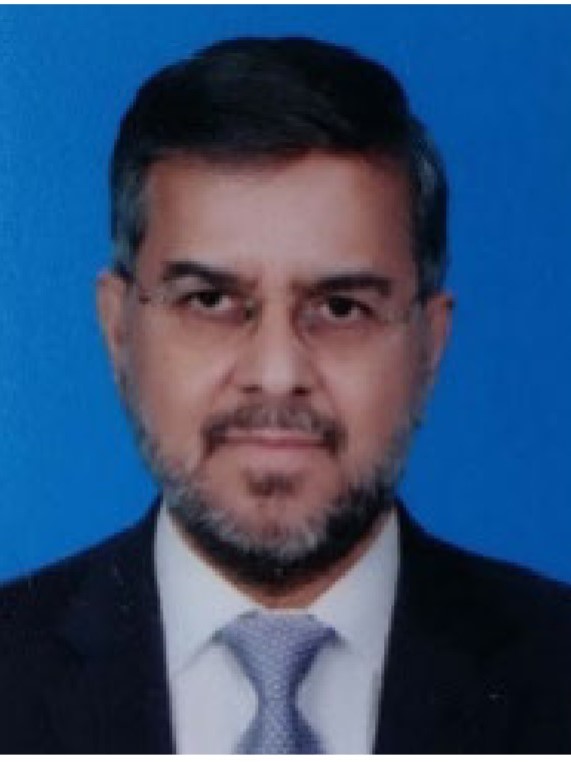}}]{Rashid Mazhar} (MBBS, MCPS, FCPS, FRCS-Glas, FRCS-Edinb.) has more than twenty years of experience in the field of cardiothoracic surgery. He is currently the head of the Thoracic surgery section at the Hamad General hospital, Doha, Qatar. Besides his surgical, clinical and educational activities, he has a particular interest and track record for collaborating with engineering colleagues to bring about user-end driven innovative solutions in his field of specialization. Automation, objective monitoring and user-friendly signal processing are his main areas of interest. Dr. Rashid is the winner of the Academic Health System Innovation award in 2014 and 2015 by proposing “Oro-Tracheal Defibrillator – a novel device to improve the outcome of CPR” and “Resuscitation Monitor - another novel device to improve the outcome of CPR”, respectively. He recently filed the patent, “Q-Stent”: which is filed through Academic Health System, HMC (PCT/IB2014/002337; WO/2016/012829). The device is already in regular clinical use at HMC paediatric and adult cardiac surgery.
\end{IEEEbiography}
\begin{IEEEbiography}[{\includegraphics[width=1in,height=1.25in,clip,keepaspectratio]{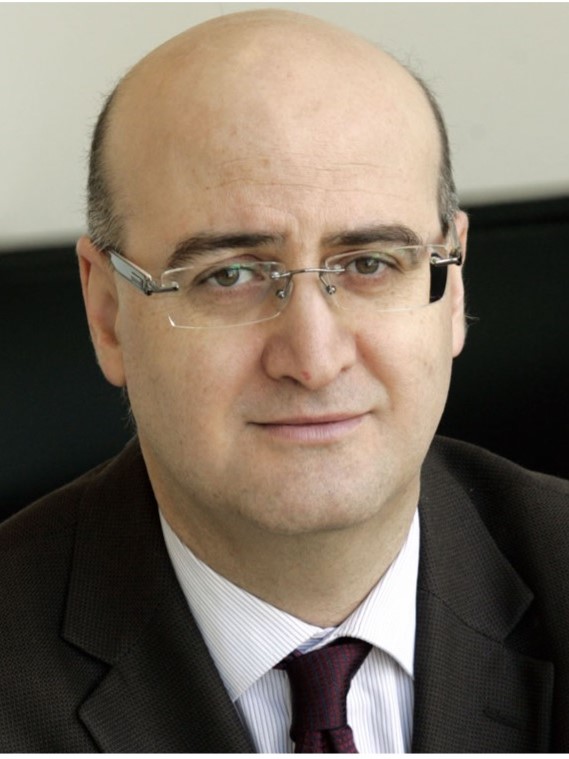}}]{Moncef Gabbouj} (F’11) is a well-established world expert in the field of image processing, and held the prestigious post of Academy of Finland Professor during 2011-2015. He has been leading the Multimedia Research Group for nearly 25 years and managed successfully a large number of projects in excess of 18M Euro. He has supervised 45 PhD theses and over 50 MSc theses. He is the author of several books and over 700 papers. His research interests include Big Data analytics, multimedia content-based analysis, indexing and retrieval, artificial intelligence, machine learning, pattern recognition, nonlinear signal and image processing and analysis, voice conversion, and video processing and coding. Dr. Gabbouj is a Fellow of the IEEE and member of the Academia Europaea and the Finnish Academy of Science and Letters. He is the past Chairman of the IEEE CAS TC on DSP and committee member of the IEEE Fourier Award for Signal Processing. He served as associate editor and guest editor of many IEEE, and international journals.
\end{IEEEbiography}

\EOD

\end{document}